\documentclass[twocolumn]{aastex7}
\usepackage{amsmath}
\usepackage{subcaption}
\usepackage{siunitx}
\usepackage{multirow}

\newcommand{\pyedith}{\texttt{pyEDITH}}

\newcommand{\snr}{$S/N$}

\begin{document}

\title{Multi-bandpass Photometry for Exoplanet Atmosphere Reconnaissance (MPEAR) with the Habitable Worlds Observatory (HWO)\\I. Differentiating Earth from Neptunes During Discovery
}

\author[orcid=0000-0002-0006-1175]{Eleonora Alei}
\altaffiliation{NASA Postdoctoral Fellow}
\affiliation{NASA Goddard Space Flight Center, 8800 Goddard Rd, Greenbelt, 20771, MD, USA}
\email[show]{eleonora.alei@nasa.gov}  

\author[orcid=0000-0002-8119-3355]{Avi M. Mandell}
\affiliation{NASA Goddard Space Flight Center, 8800 Goddard Rd, Greenbelt, 20771, MD, USA}
\email{avi.mandell@nasa.gov}  

\author[orcid=0000-0003-3429-4142]{Miles H. Currie}
\altaffiliation{NASA Postdoctoral Fellow}
\affiliation{NASA Goddard Space Flight Center, 8800 Goddard Rd, Greenbelt, 20771, MD, USA}
\email{miles.h.currie@nasa.gov}  

\author[orcid=0000-0002-2989-3725]{Aki Roberge}
\affiliation{NASA Goddard Space Flight Center, 8800 Goddard Rd, Greenbelt, 20771, MD, USA}
\email{aki.roberge-1@nasa.gov}  

\author{Christopher C. Stark}
\affiliation{NASA Goddard Space Flight Center, 8800 Goddard Rd, Greenbelt, 20771, MD, USA}
\email{christopher.c.stark@nasa.gov}  

\author{Allison Payne}
\affiliation{NASA Goddard Space Flight Center, 8800 Goddard Rd, Greenbelt, 20771, MD, USA}
\email{allypaynesmail@gmail.com}  

\author[orcid=0000-0002-5060-1993]{Vincent Kofman}
\affiliation{Centre for Planetary Habitability (PHAB), Department of Geoscience, University of Oslo, Oslo, Norway}
\email{vincent.kofman@geo.uio.no}

\author[orcid=0000-0002-2662-5776]{Geronimo L. Villanueva}
\affiliation{NASA Goddard Space Flight Center, 8800 Goddard Rd, Greenbelt, 20771, MD, USA}
\email{geronimo.villanueva@nasa.gov}  

\author[orcid=0000-0003-2215-8485]{Renyu Hu}
\affiliation{Jet Propulsion Laboratory, California Institute of Technology, Pasadena, CA 91109, USA}
\email{renyu.hu@jpl.nasa.gov}

\author[0000-0003-3099-1506]{Amber V. Young}
\affiliation{NASA Goddard Space Flight Center, 8800 Goddard Rd, Greenbelt, 20771, MD, USA}
\email{amber.v.young@nasa.gov}  

\begin{abstract}
As the architecture for the Habitable Worlds Observatory (HWO) is being developed, it is crucial to optimize the observing strategies for a survey to detect and characterize Earth-like planets around Sun-like stars. Efficient target identification and characterization will help drive mission requirements that can be matched to the planned observations. 

Current HWO concepts allow simultaneous multi-bandpass observations with the coronagraph instrument, critical for performing a qualitative planetary reconnaissance to optimize observing time for deriving orbital constraints and prioritize characterization of promising targets. 

We describe a new {open-source} algorithm designed to determine the best combination of broadband photometric observations for extracting maximum information from the first visit. It identifies degeneracies in the orbital configurations, fluxes, and noise, and determines optimal secondary photometry bands to reduce these. We demonstrate its application by comparing an Earth seen at quadrature with a cold and a warm Neptune at inclined orbits and varying phases, with comparable flux in the discovery bandpass centered at 500 nm (20\% bandwidth). 

Using the noise and exposure time calculator that we developed for the HWO coronagraph instrument, we find that the baseline $S/N=7$ (corresponding to 3.2 hours observing time for a planet at 10pc) is only sufficient to marginally differentiate the Earth from a cold Neptune-like planet assuming two parallel bandpasses (550 nm + 850 nm). However, increasing to $S/N=15$ (7 hours observing time) and using three parallel bandpasses ({360 nm }+ 500 nm + 1.11 micron) would differentiate the Earth from either a warm or cold Neptune.

\end{abstract}

\section{Introduction} \label{sec:intro}

The characterization of habitable terrestrial exoplanets is one of the main goals of the major space agencies, as highlighted in the US Astro2020 Decadal Survey \citep{NAP26141}. Astro2020 recommended the pursuit of a technical and scientific study for the Habitable Worlds Observatory (HWO), an IR/O/UV (i.e., infrared/optical/ultraviolet) “high-contrast direct imaging mission with a target off-axis inscribed diameter of approximately 6 meters” which resembles but does not coincide with the previous HabEx \citep[Habitable Exoplanet Observatory,][]{Gaudi2020}  and LUVOIR \citep[Large UV/Optical/IR Surveyor,][]{LUVOIR} concepts. In recent months, efforts have ramped up to perform architecture trade studies to eventually define an HWO concept, and the community has been heavily involved in proposing relevant science cases for the new Observatory. This is a critical task at this point in time since, as stated in the Decadal, “implementation should start in the latter part of the decade [2022-2032], with a target launch in the first half of the 2040s”, pending an independent review on the scientific and technological maturity. 
As the architecture for HWO is being developed, it is crucial to optimize the observing strategies for efficient target identification and characterization so that the mission requirements can be matched to the planned observations. 

A main exoplanet-related requirement for an HWO-like mission is to observe 30 - 50 terrestrial planets around nearby FGK stars over the span of 5 mission years, though the potential of a serviceable observatory is also discussed and would extend and enhance the nominal mission. Even with state-of-the-art coronagraphs, the spectroscopic characterization of planets whose flux is 10 billion times smaller than their host star (approximately the contrast we would need to detect an Earth-like planet around the Sun) would require days of observation \citep[see e.g.][]{2025ApJ...990...48U}. Furthermore, the presence of a coronagraph limits us to only observe a portion of the wavelength range at a time, thus increasing the required observing time for a full characterization of the planetary spectrum from the UV (i.e., ultraviolet) to the NIR (i.e. near-infrared). 

It is therefore essential to define optimal observing strategies for efficiently identifying which planetary discoveries are most likely to be habitable, and therefore of highest priority for spectroscopic follow-up. For the best targets, HWO will acquire spectroscopic data on the full wavelength range, and the current strategy baselined for target prioritization is through spectroscopy ``decision tree'' observation sequences \citep{Young2024} which prioritize key molecules to detect that can distinguish different planet types. Bayesian retrievals have been used to identify what is the minimum and optimal wavelength ranges to perform observations for a variety of planetary atmospheres \citep[see e.g.,][]{2018AJ....155..200F,2022AJ....163..299D,2023AJ....166..157D, 2023AJ....166..129L, 2025arXiv250714771K,2025ApJ...990...48U}; these observational requirements have recently been used when simulating exo-Earth survey yields \citep{2024AAS...24324906M,2024SPIE13092E..5MM, Stark2024, Stark2}. 

To date, the potential for acquiring preliminary planetary composition information during the photometric planet detection phase has not been explored in depth for HWO. The detection phase, currently considered to be a "blind" survey of high-priority stellar targets, is a step that precedes spectral characterization and that is necessary to identify new planets to be studied. Follow-up photometric observations would then be required to determine the orbit of the candidate, and only then a prioritization would be made to identify the targets to follow up with spectroscopy. However, since the mission time allowed for the large-scale exoEarth discovery survey is limited, it would be sensible to perform orbital mapping and spectral characterization of only the highest-priority planetary discoveries. For this reason, the prioritization should happen as early in the survey process as possible. 

Due to detector performance limitations, HWO is currently expected to have three separate wavelength channels: one in the ultraviolet (UV), from $\sim$0.2 to $\sim$0.4 $\mu$m, one in the visible (VIS), from $\sim0.4$ to $\sim1.0$ $\mu$m, and one in the near-infrared (NIR) range, from $\sim$1.0 to $\sim1.8$ $\mu$m. The long and the short wavelength cutoffs are still subject to discussion \citep{2025arXiv250714771K}, as well as the boundaries between ``parallel" (simultaneous) channels.
One current strategy is that HWO would acquire a broadband photometric observation at 500 nm during the first visit \citep{Stark2024}, but there is potential for multiple parallel channels covering separate bandpasses.  Acquiring multiple, simultaneous broadband photometric observations during the first visit could enable us to: 1) photometrically identify and recognize different detected planets when observing multi-planet systems, in order to optimize the orbit-fitting process over multiple visits; and 2) collect key information to differentiate habitable and uninhabitable planets from the first visit and prioritize follow-up observations. This approach enables a more efficient use of telescope resources and increases the likelihood of identifying promising candidates for detailed study. 

The potential of colors and broadband photometry to qualitatively characterize exoplanets is not new. \citet{2003ASPC..294..595T} argued that colors and low-resolution spectra in the VIS range would allow us to characterize atmospheric and surface spectral signatures, as well as determine physical parameters. 
On the other hand, \citet{Cahoy2010} performed a study on the albedo of gas giants and highlighted that phase-dependence and the presence of clouds could make this less straightforward and multiple observations would be needed to determine cloud distributions and orbital parameters. \citet{JKT2016} empirically calculated the optimal photometric bins to differentiate an Earth from a variety of other planetary classes. To achieve this, they identified the filters that maximized the distance in color-color space between the Earth and every other modeled planet. \citet{2018AJ....156..158B} explored the classification of gas giants through color information obtained with potential filter options for the Nancy Grace Roman Space Telescope, and the relation with metallicity, temperature, clouds, and phases. The authors performed a supervised classification multivariate analysis to assess if photometry would be sufficient to retrieve the underlying physics of an exoplanet. The analysis was done every combination of two up to five filters and found that at least three filters are needed to provide any classification.
\citet{Smith2020} calculated a grid of potential rocky atmospheres and evaluated the potential for multiband photometry to effectively distinguish different classes of planetary atmospheres. They found the technique effective for atmospheres exhibiting strong absorption lines, but more challenging for Earth-like abundances and requiring large apertures to improve results.

Beyond the challenges of degeneracies in planetary atmospheres, distinguishing an Earth from other planetary types is made even more difficult when considering orbital geometry effects and unknown phase. As shown in \citet{2018haex.bookE..98R}, the reflected light spectra of different planets span orders of magnitude when positioned at the same semi-major axis and phase. However, degeneracies can appear when altering these two parameters, and atmospheric features might become less distinguishable. Specifically, \citet{2018haex.bookE..98R} highlighted a scenario where a Neptune-like planet and an Earth would hardly be differentiable with each other at a given phase, especially at lower resolutions and within the average expected noise for current instrumentation. Given the limited observing time available for future direct imaging missions, developing strategies to efficiently distinguish between these planetary types becomes crucial for prioritizing truly habitable worlds over larger, uninhabitable planets that happen to fall within similar brightness ranges. 

Building on this work, we embarked on a study to assess the potential of multi-bandpass photometry in differentiating the Earth from two template Neptune-like planets with HWO. This serves as a preliminary step that would allow us to identify potentially promising targets to follow up with orbit determination and spectroscopic characterization. This first paper is a proof-of-concept of a methodology that could then be expanded to encompass multiple planetary classes and observing strategies.

We provide methodology and results for every step in the algorithm: Section \ref{sec:orbits} describes the orbital modeling; Section \ref{sec:fluxes} deals with the radiative transfer flux calculations; Section \ref{sec:noise} provides details on the Exposure Time Calculator for HWO and the noise simulations; Section \ref{sec:observations} explores the best filter combinations for a successful discrimination between an Earth and Neptune-like planets. We discuss the implications and limitations of this study in Section \ref{sec:discussion}, and we provide a summary in Section \ref{sec:summary}.

\section{Methodology Overview} \label{sec:methods}

We start with the scenario of a detection of an assumed planetary object at a given separation from the star, with an orbital separation and flux at 500 nm compatible with an Earth twin on a face-on orbit. However, other orbital configurations and planetary properties may also be compatible with these minimal constraints, causing ``planetary confusion". Our goal is to understand what these compatible scenarios are, and how the observing strategy can be optimized to break these degeneracies in planetary properties and correctly prioritize promising habitable planets with respect to uninhabitable ones. We consider a planet detection to be ``confusing" if two conditions are true: 1) It is reasonable to find the planet in the detected position at any given time (i.e., if the orbit crosses the detected position); 2) The photometric flux at 500 nm (i.e., the detection band currently assumed for HWO) yields compatible results with what the flux of an Earth twin in that configuration would be. To achieve this, we create an algorithm that models orbits, fluxes, and noise simulations for given planetary classes, and then we assess what combination of simultaneous photometric observations are preferable for disentangling the possible scenarios. 

We test this algorithm considering the Earth on a face-on orbit (quadrature phase) compared to two template Neptune scenarios (which we call ``Warm" and ``Cold"). We will expand on different phases and more realistic scenarios in an upcoming study (see Section \ref{sec:limitations}).

\section{Orbital Calculations} \label{sec:orbits}

The projected distance for an Earth-twin orbiting at 1 AU from a Sun-{twin} star at a distance of 10 pc on a face-on orbit ($i=0^\circ$) would be 0.1 arcseconds. Our first task is to identify all orbital configurations that would also satisfy the first required condition, i.e. that the planet is on an orbit that allows it to be detected at a given projected separation from the star. To do this, we need to calculate all orbits that are consistent with the specified projected distance. We implement a comprehensive search across the orbital parameter space, calculating millions of planetary positions in possible orbits and identifying those that place the planet at our target coordinates.

We implemented an algorithm that solves the Keplerian orbital calculation, following \citet{2010exop.book...15M}. We assume that the orbit lies on a  $(\hat{x}, \hat{y}, \hat{z})$ coordinate system, and that the observer is in the reference frame  $(\hat{X}, \hat{Y}, \hat{Z})$ \citep[see Figure 4 of][]{2010exop.book...15M}. Three angles define the inclination of the orbit with respect to the reference frame: the inclination $i$, defined as the angle between the orbital plane and the reference plane, which dictates the tilt of the orbit in the ($\hat{Y}, \hat{Z}$) plane;  the longitude of the ascending node $\Omega$, which determines the orientation of the orbit in the ($\hat{X}, \hat{Y}$) plane; and the argument of periapse $\omega$, which determines where the periapse of the orbit is located with respect to the radius vector of the ascending node.

We define $r$ as the radius of the orbit (whose value depends on the semi-major axis of the orbit $a$ and the eccentricity $e$) and $f$ as the true anomaly, the angle between the radius vector of the periapse and the radius vector of the orbital position of the planet. In the  $(\hat{x}, \hat{y}, \hat{z})$ coordinate system, the planet would have coordinates equal to $x=r\cos(f)$, $y=r\sin(f)$, and $z=0$. In the $(\hat{X}, \hat{Y}, \hat{Z})$ reference system, these translate into:
\begin{equation}
    \begin{cases}
        X = r  [\cos\Omega \cos(\omega+ f) - \sin\Omega \sin(\omega + f) \cos i]\\
    Y = r[ \sin\Omega \cos(\omega + f) +\cos\Omega \sin(\omega + f) \cos i] \\
    Z = r \sin(\omega + f) \sin i
    \end{cases}
\end{equation}

We can calculate the phase of the planet with respect to the observer by calculating the angle between the planet-star and the planet-observer vectors. The first vector (planet-star) is the negation of the one calculated previously:

\begin{equation}
    \vec{ps} = [-X, -Y, -Z]
\end{equation}

And the second vector (planet-observer) is the difference between the vector connecting the star to the observer (assuming a distance $D$) and the vector connecting the star to the planet. This translates into:
\begin{equation}
    \vec{po} = [0,0,D]-[X, Y, Z] = [-X,-Y,D-Z]
\end{equation}

The angle between the two vectors is defined as:
\begin{equation}
    \phi = \arccos\left( \frac{\vec{po} \cdot \vec{ps}}{ |\vec{po}| |\vec{ps}|}\right)
\end{equation}

In our calculations we considered circular orbits ($e=0$, $r=a$). Furthermore, we made the following assumptions:
\begin{itemize}
    \item $\Omega=0^\circ$. This sets the line of nodes to lie on the \emph{x} axis and reduces the numbers of models to simulate. This is acceptable for our study since we focus on single-epoch photometry: the phase angle of the planet at one single epoch is independent on the orientation of the orbit. This parameter will become important when analyzing multi-epoch observations, where the orbit determination becomes essential. 
    \item $\omega = -90^\circ$. The argument of the periapse cannot be defined for circular orbits. We select a specific angle to allow the common convention used in exoplanet research to define $\phi=180^\circ$ for a transiting planet ($i=90^\circ$). This also defines $t=180^\circ$ in this specific scenario.  
    \item $i$ is conventionally defined as the angle between the (positive) $(\hat{X},\hat{Y})$ and the $(\hat{x},\hat{y})$ planes. Since the target position on the $(\hat{X},\hat{Y})$ plane (i.e., the field of view), is located at negative values of $Y$, the inclinations that would match the target would have $90^\circ\le i \le 180^\circ$, which is counter-intuitive. We therefore define $\bar{i}$ as the supplementary angle of $i$ as the variable under our control, and the algorithm translates the angles internally. In terms of phase angle, this has no impact since the geometry of the single epoch is symmetric with respect to the y axis (i.e. the illuminated portion of the planet will be the same whether the planet lies on positive or negative values of Y, provided the Z coordinate is the same).
\end{itemize}

We generate orbits for every combination of the following three parameters: 1) semi-major axis $a$, allowed to vary between 1 and 35 AU at steps of 0.25 AU; 2) inclination $\bar{i}$ (see above), allowed to vary between 0$^\circ$ and 180$^\circ$ at steps of 1$^\circ$; 3) true anomaly $f$, allowed to vary between 0$^\circ$ and 360$^\circ$ at steps of 1$^\circ$. This sums to $\sim$25K orbits and 8.8 million orbital positions to be tested.

\subsection{Results}

\begin{figure*}
\centering
\begin{subfigure}[b]{\textwidth}
    \centering
    \includegraphics[width=0.74\linewidth]{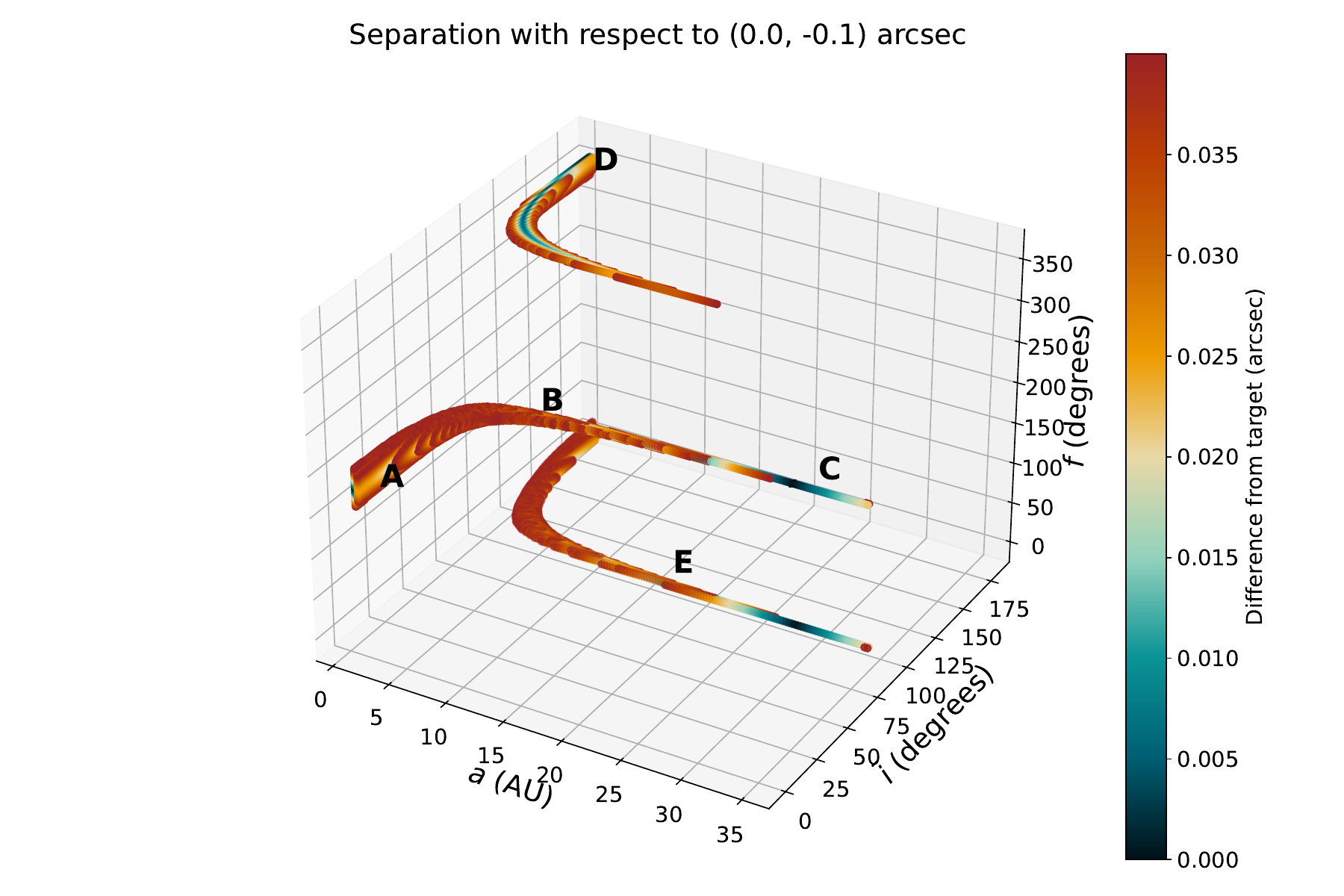}
\end{subfigure}

\begin{subfigure}[b]{0.47\linewidth}
    \centering
    \includegraphics[width=\textwidth]{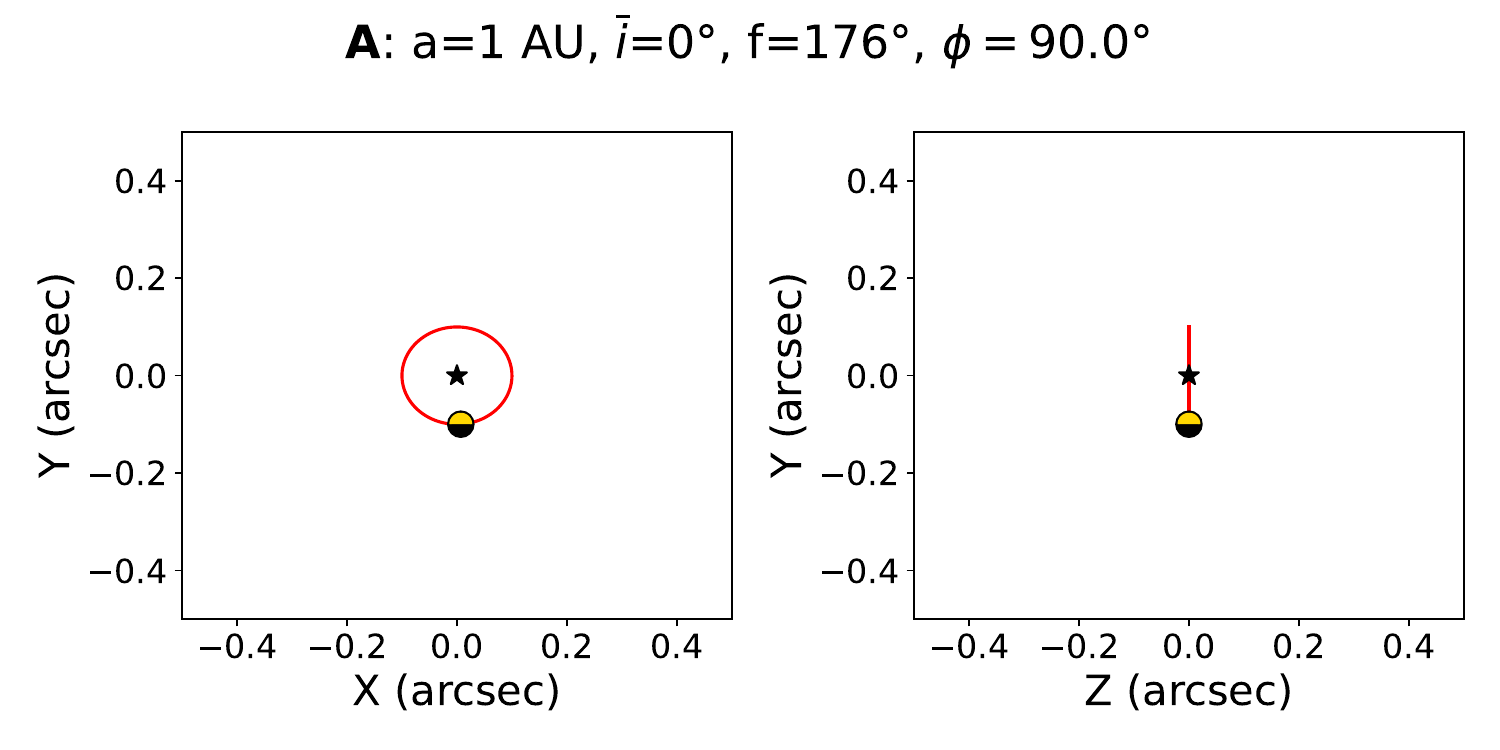}
\end{subfigure}
\hfill
\begin{subfigure}[b]{0.47\linewidth}
    \centering
    \includegraphics[width=\textwidth]{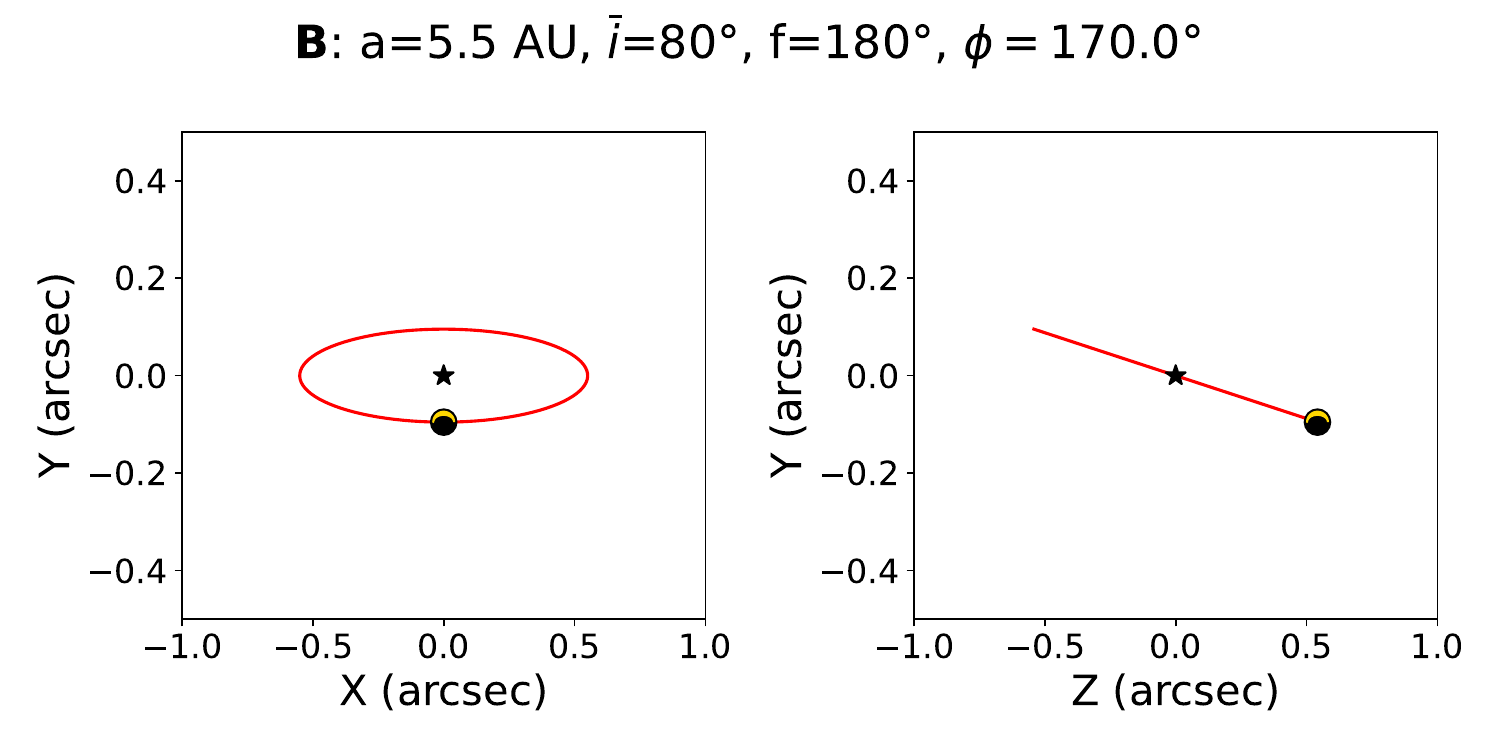}

\end{subfigure}

\begin{subfigure}[b]{0.47\linewidth}
    \centering
    \includegraphics[width=\textwidth]{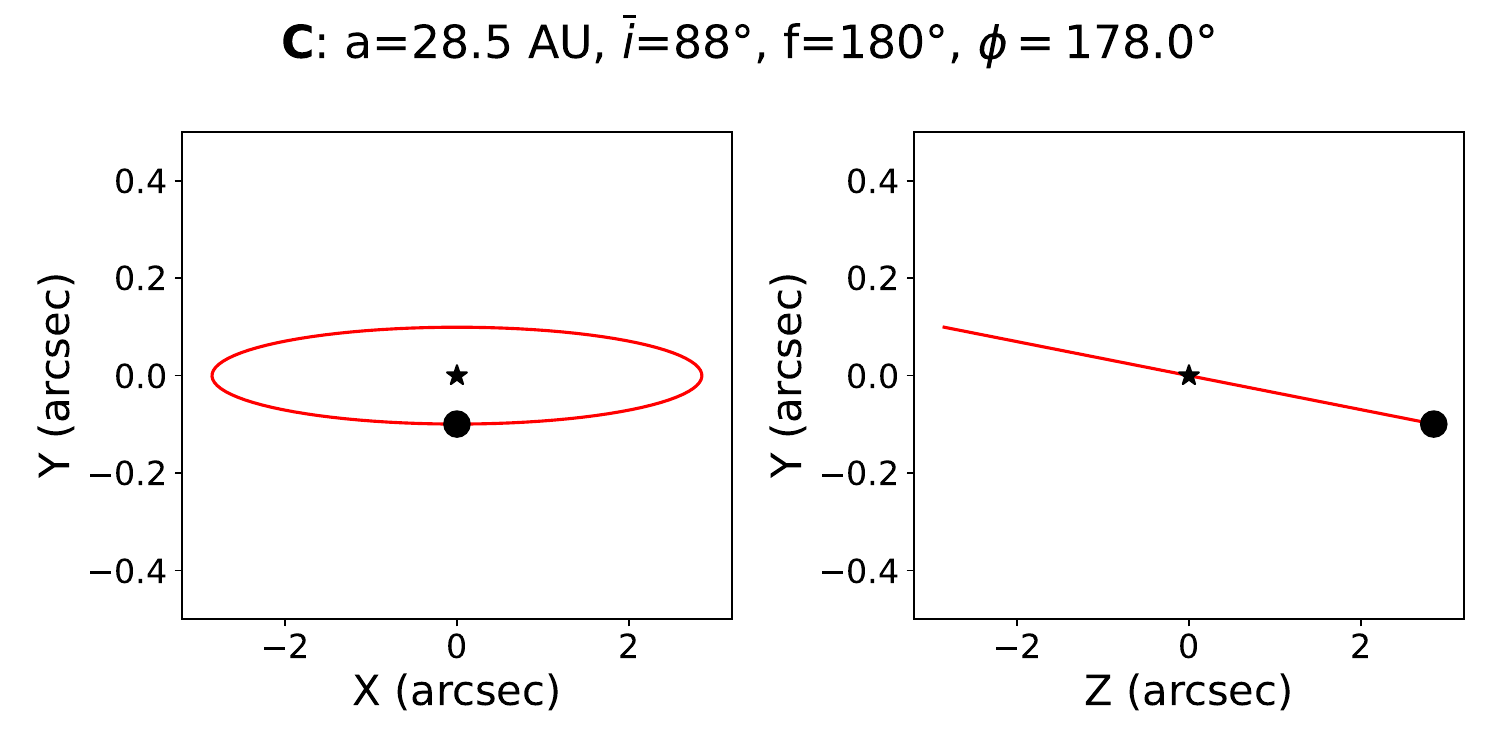}

\end{subfigure}
\hfill
\begin{subfigure}[b]{0.47\linewidth}
    \centering
    \includegraphics[width=\textwidth]{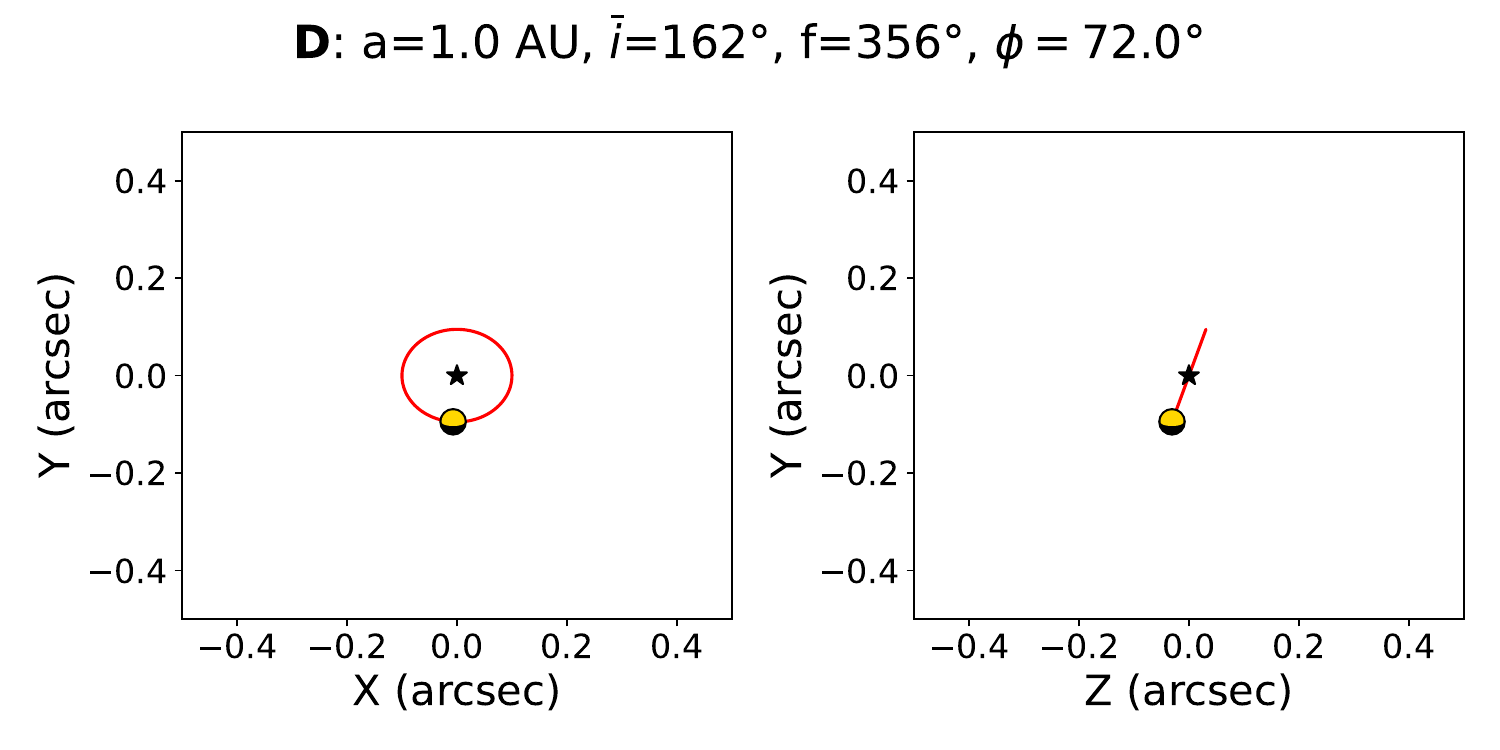}

\end{subfigure}

\begin{subfigure}[b]{0.47\linewidth}
    \centering
    \includegraphics[width=\textwidth]{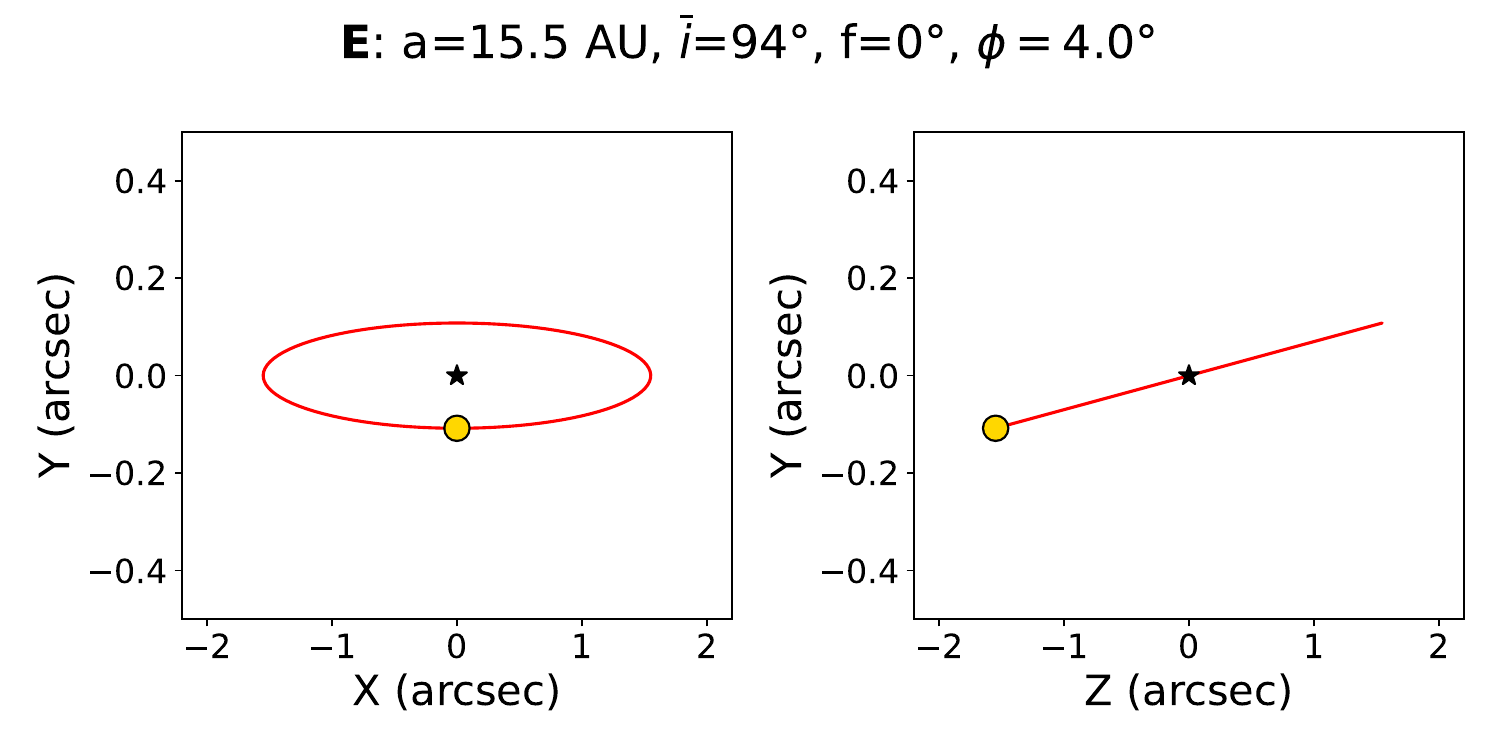}

\end{subfigure}
\hfill
\begin{subfigure}[b]{0.47\textwidth}
    \centering
    \includegraphics[width=\textwidth]{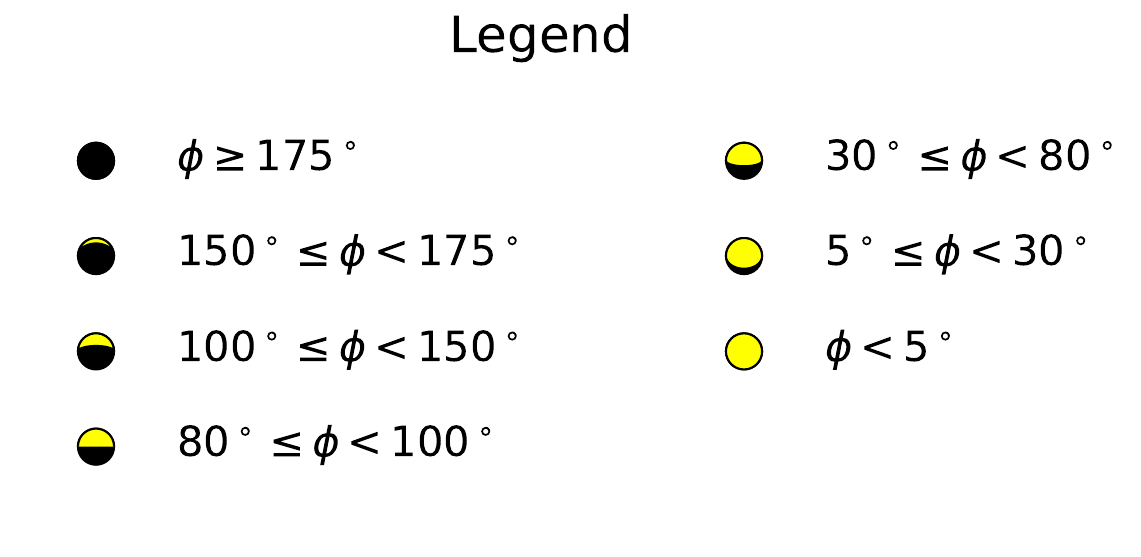}
\end{subfigure}
\caption{Some combinations of semi-major axes, inclination, and true anomaly that were considered for this study. \emph{Top panel}: 3D scatter plot showing the combinations of parameters that would yield a plausible orbit that would match the target at (0,-0.1) arcsec. The colorbar shows the absolute difference of the coordinates from the target position. Configurations are considered to be ``compatible" with the target position if the difference is less than 0.01 arcsec. \emph{Bottom panels}: Individual orbit plots for specific points A through E labeled in the top panel, each with their projection in $\hat{X}-\hat{Y}$ and $\hat{Z}-\hat{Y}$ space. For each point, details on the orbital location in terms of semi-major axis, inclination, true anomaly and apparent phase are shown. In the various plots, special markers indicate visually the phase that we would expect to find the planet in.}
\label{fig:3dscatterorbits}
\end{figure*}

In the top panel of Figure \ref{fig:3dscatterorbits} we show in a 3D plot the combinations of semi-major axis $a$, inclination $\bar{i}$, and true anomaly $f$ that yield orbits compatible with the assumed detected planet. The color bar represents the two-dimensional difference from the target position in ($\hat{X}-\hat{Y}$) space (i.e., the field of view). In this plot, we only show combinations that yield a difference lower than 0.04 arcsec for clarity. In the bottom five panels we show the orbits of five points identified in the 3D plot for illustrative purposes, with markers showing the expected phase of the planet at the detection point. These are shown in both ($\hat{X}-\hat{Y}$) and ($\hat{Z}-\hat{Y}$) projection. The observer would be located at positive values of $\hat{Z}$.

Intuitively, the target position that is (by construction) consistent with a face-on orbit of an Earth-like planet at quadrature (point A in Figure \ref{fig:3dscatterorbits}) can also be consistent with a larger orbit almost edge-on (point C in Figure \ref{fig:3dscatterorbits}). At shorter semi-major axes, orbits almost face-on ($\bar{i}\approx0^\circ$ or $\bar{i}\approx180^\circ$) with true anomalies around $0^\circ$ or $180^\circ$ are more likely to yield a small difference. These configurations are very similar: one slightly inclined towards the observer (positive $\hat{Z}$ axis), one further from it (negative  $\hat{Z}$ axis), which determines the different denomination of the true anomaly (see e.g., Point A and Point D).  Such orbits are relatively similar to the Earth in quadrature scenario.  

At increasingly larger semi-major axes, only orbits close to edge-on can be compatible ($\bar{i}\approx90^\circ$), with true anomalies very close to $180^\circ$ and $0^\circ$ (depending on whether their $\hat{Z}$ coordinate is positive or negative, see e.g., points E and C).

With the orbital calculations we performed, we can determine what subset of semi-major axes, inclinations, and true anomalies (with their derived planetary phases) are compatible with the detected target position (satisfying the first condition).
 We assume the condition to be satisfied if the difference from the target position is lower than 0.01 arcsec (10\% of the projected separation).
 We find that 1435 models out of 8.8 million orbital configurations (corresponding to 409 out of 25,000 unique orbits) satisfied the orbital condition. This means that, when observing the astrophysical scene during the first visit, with no information on the orbit of the planet, any of the 1435 orbital configurations could be the true scenario. 
 
 \begin{figure}
     \centering
     \includegraphics[width=\linewidth]{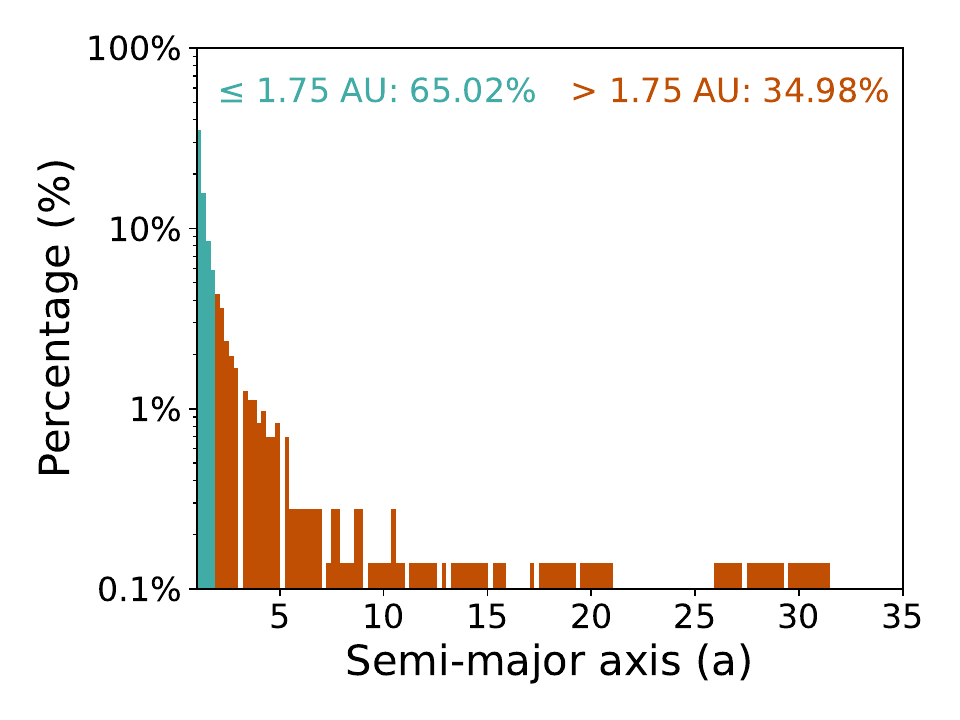}
     \caption{Histogram showing the percentage of orbits that satisfy the orbital condition with respect to the semi-major axis parameter. In green, the orbits whose semi-major axis is $\le1.75$ AU; in red, the orbits with $a > 1.75$ AU. Labeled at the top, the cumulative percentages of all orbits belonging to these subgroups. Gaps in the results are caused by the discrete sampling of the grid.}
     \label{fig:semimajoraxis}
 \end{figure}

 In Figure \ref{fig:semimajoraxis} we show how orbits that satisfy condition 1) are distributed in semi-major axis space. We highlight in green the models that are {within} the commonly assumed boundaries for the habitable zone \citep[HZ; ][]{Kopparapu2013-HABITABLEZONESMAINSEQUENCE}, and in red are the ones outside the HZ {(note that planets interior to the HZ would have smaller angular separations than our assumed 0.1 arcsec and thus would not satisfy the orbital condition)}.  We notice that about 65\% of the orbits that fit the assumed target projected separation are within the HZ. A non-negligible 35\% of the compatible orbits lie well beyond the HZ, with most within 5 AU. If we assume no prior knowledge on the observed system, we would have no way of knowing from the first orbit which of these configurations is the correct one. While we could assume that it is more likely that a planet at the prescribed 0.1 arcsec separation is explained by an orbit within the HZ, scenarios including planets outside the HZ cannot be excluded.
 
It must be noted that these results are highly dependent on the uncertainty of the projected separation. For this study, we made an arbitrary choice to ensure that the potential Earth would be located within the HZ of its star (which explains the higher percentage of planets in this region), but in reality these requirements might be more relaxed, especially for more distant targets. We also assumed that a planet could be potentially found at any inclination or semi-major axis, disregarding any information that could come from planet population occurrence rates or precursor observations. We address the limitations of these assumptions in Section \ref{sec:limitations}.

\section{Flux Calculation}\label{sec:fluxes}

Up to this point, our calculations have not taken into account any potential information that will be gathered on the flux of the planet during the first visit. The next step is to identify which configurations also satisfy the second condition: the models yield a photometric flux at 500 nm compatible with the one provided by an Earth twin at 1 AU distance on a face-on orbit. To do this, we first calculate a reference flux of the Earth at quadrature. Then, we calculate Neptune-like spectra for each combination of semi-major axes, inclinations, and true anomalies previously identified, and then compare which subset of these also satisfies the second condition.

\begin{table*}
\caption{Details of the templates for Earth and the two Neptune scenarios used in this work. $^a$: we consider the reference level either the surface (Earth) or the level at which the atmosphere is optically thick (Neptunes).}
\label{tab:templates}
\centering
\begin{tabular}{p{4.9cm}p{4cm}p{4cm}p{4cm}}
\hline
Parameter & Earth & Cold Neptune & Warm Neptune \\
\hline\hline
Object Diameter [km] & 12742 & 49248 & 49248\\
Gravity $[\mathrm{m\ s^{-2}}]$ & 9.81 & 11.28 & 10\\
Mean Molecular Weight & 28.97 & 2.6 & 2.6 \\
Bulk Gases & N\textsubscript{2}, O\textsubscript{2} & H\textsubscript{2}, He & H\textsubscript{2}, He \\
Trace Gases & H\textsubscript{2}O, CO\textsubscript{2}, O\textsubscript{3}, N\textsubscript{2}O, CO, CH\textsubscript{4}   & CH\textsubscript{4},  CO, H\textsubscript{2}O, HCN, NH\textsubscript{4}SH & H\textsubscript{2}O, NH\textsubscript{3}, CH\textsubscript{4}, H\textsubscript{2}S\\

Clouds & Water (ice), Water (liquid) & Ammonia (ice), Water (ice) & Ammonia, Water (ice) \\
Scattering Processes  & Rayleigh, Refraction, CIA & Rayleigh, Refraction, CIA, Raman & Rayleigh, Refraction, CIA, Raman\\

Pressure at reference level$^a$ [bar] & 1 & 100 & 100 \\
Temperature at reference level$^a$  [K] & 300 & 220 & 804 \\
Surface Albedo & 0.39 (66\% ocean, 3\% snow, 15\% grass, 8\% soil, 8\% forest) & N/A & N/A \\
Surface Scattering & Lambertian & N/A & N/A\\
Atmosphere Reference & \citet{kofman24} & \citet{2025arXiv250813368P} & \citet{2019ApJ...887..166H}\\
\hline
\end{tabular}

\end{table*}

\begin{figure*}
    
        \includegraphics[width=\linewidth]{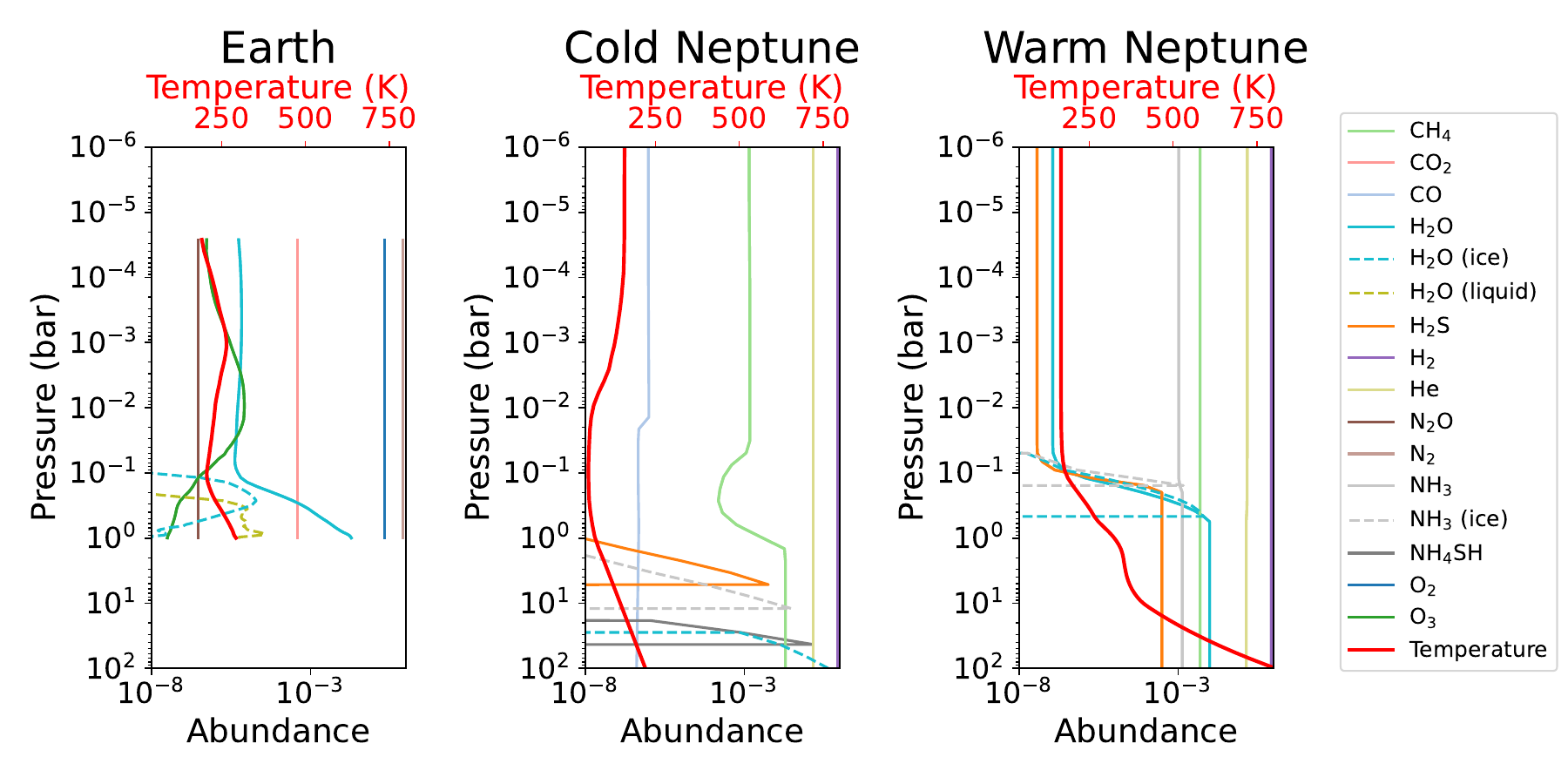}

    \caption{Temperature and abundance profiles for template Earth and Neptunes atmospheres. \emph{Left:} Earth atmosphere from \citet{kofman24} \emph{Center:} Cold Neptune atmosphere from \citet{2025arXiv250813368P} \emph{Right:} Warm Neptune atmosphere from \citet{2019ApJ...887..166H}.}
\label{fig:templates}
\end{figure*}

To calculate planetary reflectance spectra, we use the Planetary Spectrum Generator (PSG)\footnote{\url{https://psg.gsfc.nasa.gov}} \citep{Villanueva2018}.  
 PSG is a radiative transfer model suite hosting tools that are widely used to model not only confirmed exoplanets, but also Solar System planets and cometary objects. PSG can compute 1D and 3D radiative transfer models from any configuration and allows the flexibility to choose instrumental parameters for nearly any observatory. 
It can ingest billions of spectral lines of almost 1000 species across its spectral repositories (HITRAN, JPL, CDMS, GSFC-Fluor, ExoMol), allowing the user to produce detailed high resolution spectra. 

When calculating planetary flux, PSG will by default calculate an average incidence and outgoing angle, thus producing an average disk-integrated flux to enable comparison with simpler 1D radiative transfer codes. However,  this treatment is not accurate enough for our purposes, since we want to correctly model the reflection spectrum of a planet given a specific phase.

We therefore enabled the ``disk sub-sampling'' mode in PSG to produce planetary reflectances at different phase angles. In this mode, PSG divides the disk of the planet into multiple concentric regions of common incidence and emission angles, each with a specific weight based on the disk percentage the region occupies. For every bin, the radiative transfer equation is solved and the output flux is a weighted mean of the fluxes of each region. In our study, we assumed a sampling of $N=5$, which creates five bins for the incidence angle and five more for the emission angle. As these two angles change with phase, the total amount of regions that are created (as the possible combinations of the incidence and emission angle) varies, up to around 25 ($\approx N^2$). \citet{2021AJ....162...30S} discusses PSG's disk sub-sampling algorithm in more depth.

The algorithm we developed for this study is interfaced with an instance of PSG installed locally via a Docker container\footnote{\url{https://docs.docker.com/get-started/docker-concepts/the-basics/what-is-a-container/}}, enabling repeated and rapid calls to PSG \citep[see][for more information on how to implement a ``dockerized'' PSG version]{Villanueva2018,2022fpsg.book.....V}. For every model to be calculated, the algorithm produces a configuration file  that describes the desired input (including the atmospheric profile, the desired opacities to be used, the surface and aerosol scattering parameters, the stellar class and the orbital parameters) as well as the format of the desired output, and then reads in the results of the calculation. 

For the reference scenario (Earth at quadrature), we calculate the flux of an Earth twin orbiting a Sun-{twin} star at 1 AU. For the atmospheric profile for a standard Earth-like planet, we use an atmosphere based on \citet{kofman24}, modeled using the Modern-Era Retrospective analysis for Research and Applications (MERRA-2) 3D climate database\footnote{\url{https://gmao.gsfc.nasa.gov/gmao-products/merra-2/}}.  The 3D distribution of temperature, pressure, clouds and relevant molecules from MERRA-2 were ingested into the Global Emission Spectrum \citep[GlobES,][]{2025A&C....5300982F} module of the Planet Spectrum Generator \citep[PSG][]{Villanueva2018}). In the MERRA-2 reanalysis database, the 3D abundance and presence of clouds was parameterized in a 2D cloud fraction field. PSG evaluates the spectra from the MERRA-2 input by running two spectral calculations, one considering a clear sky and one assuming a fully clouded atmosphere according to the ingested cloud particle abundances. The two spectra are combined following the cloud fraction. 

Surface coverage was adopted from the Moderate Resolution Imaging Spectroradiometer (MODIS)\footnote{\url{https://modis.gsfc.nasa.gov}}  using five different surface reflectances from the United States Geological Survey (USGS)\footnote{\url{https://www.usgs.gov/labs/spectroscopy-lab/usgs-spectral-library}}. The data from this source were validated with narrow band photometric observations from DSCOVR (between 0.318-0.780 microns), as well as several other observations of Earth \citep{2025arXiv250813368P}. We show in Figure \ref{fig:templates} and Table \ref{tab:templates} the atmospheric and temperature profiles for the Earth, as well as some relevant input parameters.

We then calculate the flux of all compatible orbit configurations from Section \ref{sec:orbits}, assuming two different Neptune scenarios: if the semi-major axis is less than 3.5 AU (1167 out of 1435 Neptunes), then we assume a warm Neptune atmosphere; otherwise (268 out of 1435 Neptunes), we assume a cold Neptune atmosphere. The choice of the 3.5 AU threshold is motivated by estimates for the snow line \citep[see e.g.,]{Martin2012}, which is a sensible physical separation between classes of Neptune-like planets with different atmospheric compositions and thermal structures. Beyond this boundary, a water or ammonia cloud deck is expected above $\sim$1 bar \citep{2019ApJ...887..166H, Cahoy2010}.

We use fixed atmospheric templates for each Neptune type to isolate the effects of orbital geometry (phase angle and distance) on observed fluxes, allowing us to focus on how orbital parameters alone can create spectral confusion. {We discuss the implications of this assumption in Section \ref{sec:limitations} and we will consider self-consistent chemistry in a future work.} 

Figure \ref{fig:templates} and Table \ref{tab:templates} show the the atmospheric and temperature profiles for the two Neptunes and their relevant input parameters.

As warm Neptune template, we use one of the models calculated in \citet{2019ApJ...887..166H}. We selected a model for an H$_2$/He-dominated warm Neptune at a semi-major axis 1.7 AU (a good medium between the edges of the warm Neptune subgrid), internal heat flux $T_{int}=60\ K$ to mimic a Neptune-like planet, gravity of $10\ m/s^2$ (close to that of Neptune), and 10x solar metallicity (similar to the metallicity found in observed warm Neptune-sized planets). The models were computed using ExoREL, which calculates pressure-temperature profiles using a grey-atmosphere approximation and taking into account the condensation of water and ammonia clouds in a self-consistent way. 

The cold Neptune template was derived from all available measurements of the Solar System's Neptune \citep{2025arXiv250813368P}. The cold Neptune continuum is based on data from the Hubble Space Telescope \citep{Courtin1999}, the European Southern Observatory \citep{Karkoschka1998} and the NASA Infrared Telescope Facility \citep{Irwin2022}. The collection of observations provide the composite reflectance spectrum that was used as a reference to develop the best-fit reflectance model in PSG by adjusting atmospheric absorbers, aerosols, and cloud features to match the reference spectrum; additional details are available in \citet{2025arXiv250813368P}. 

We simulate the ``discovery" flux measurement by calculating the average photometric flux of the Earth at 500 nm assuming 20\% bandwidth $F_{500,20\%}$. This translates into averaging the flux in the interval $[500\cdot(1-10\%),500\cdot(1+10\%)]$ nm $= [450,550]$ nm. Since the currently-assumed signal-to-noise ratio (\snr) for detection is 7 \citep{Nemati2020-MethodDerivingOptical} at the detection wavelength range, the errorbar associated with the measurement $\Delta F_{500,20\%}=F_{500,20\%}/7$. The flux of a Neptune is then considered compatible if the averaged Neptune flux over the discovery bandpass lies within $F_{500,20\%}\pm\Delta F_{500,20\%}$. 

\subsection{Results}

\begin{figure*}
    \centering
    \includegraphics[width=\linewidth]{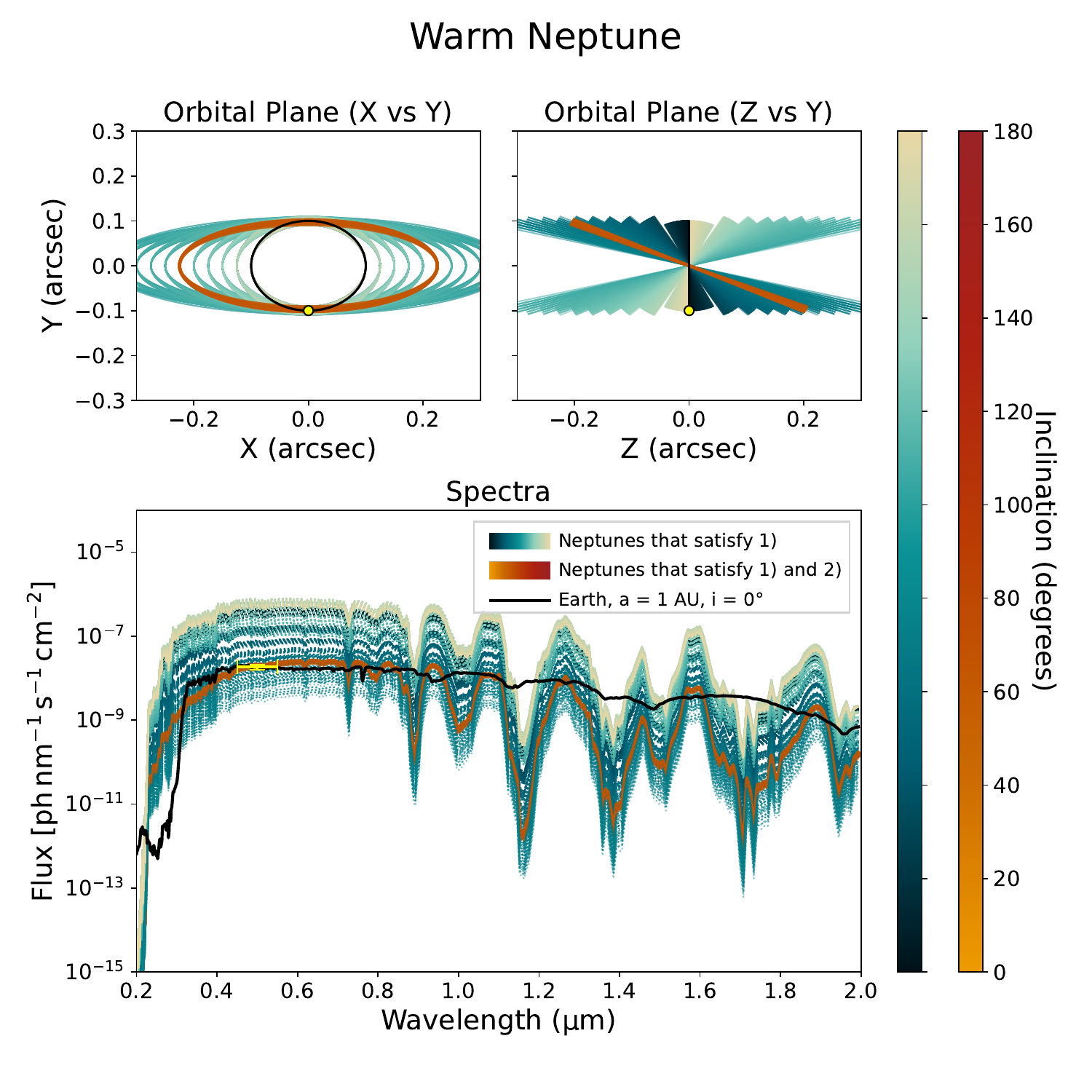}
    \caption{Orbits (\emph{top left panel}: $\hat{X}-\hat{Y}$ orbital projection; \emph{top right panel}: $\hat{Z}-\hat{Y}$ orbital projection) and fluxes (\emph{bottom panel}) of the warm Neptunes that are compatible with the assumed detection point, compared to the orbit and flux of an Earth at quadrature. In black, the orbit (top panels) and the flux (bottom panel) of the Earth on a face-on orbit. In yellow-green hues, the orbits (top panels) and fluxes (bottom panels) of all warm Neptunes that are compatible with the detected position (Condition 1), color-coded according to the inclination of their orbits.  In red hues, the orbits (top panels) and fluxes (bottom panel) of the warm Neptunes that also yield a comparable flux at 500 nm (Condition 2). In the top panels, the projected distance is shown as a yellow circle. In the bottom panel, the photometric spectrum of the Earth template calculated at 0.5 $\mu$m with a 20\% bandwidth (horizontal errorbar), and assuming \snr=7 (vertical errorbar), is shown in yellow. }
    \label{fig:warmneptuneorbitsspectra}
\end{figure*}

\begin{figure*}
    \centering
    \includegraphics[width=\linewidth]{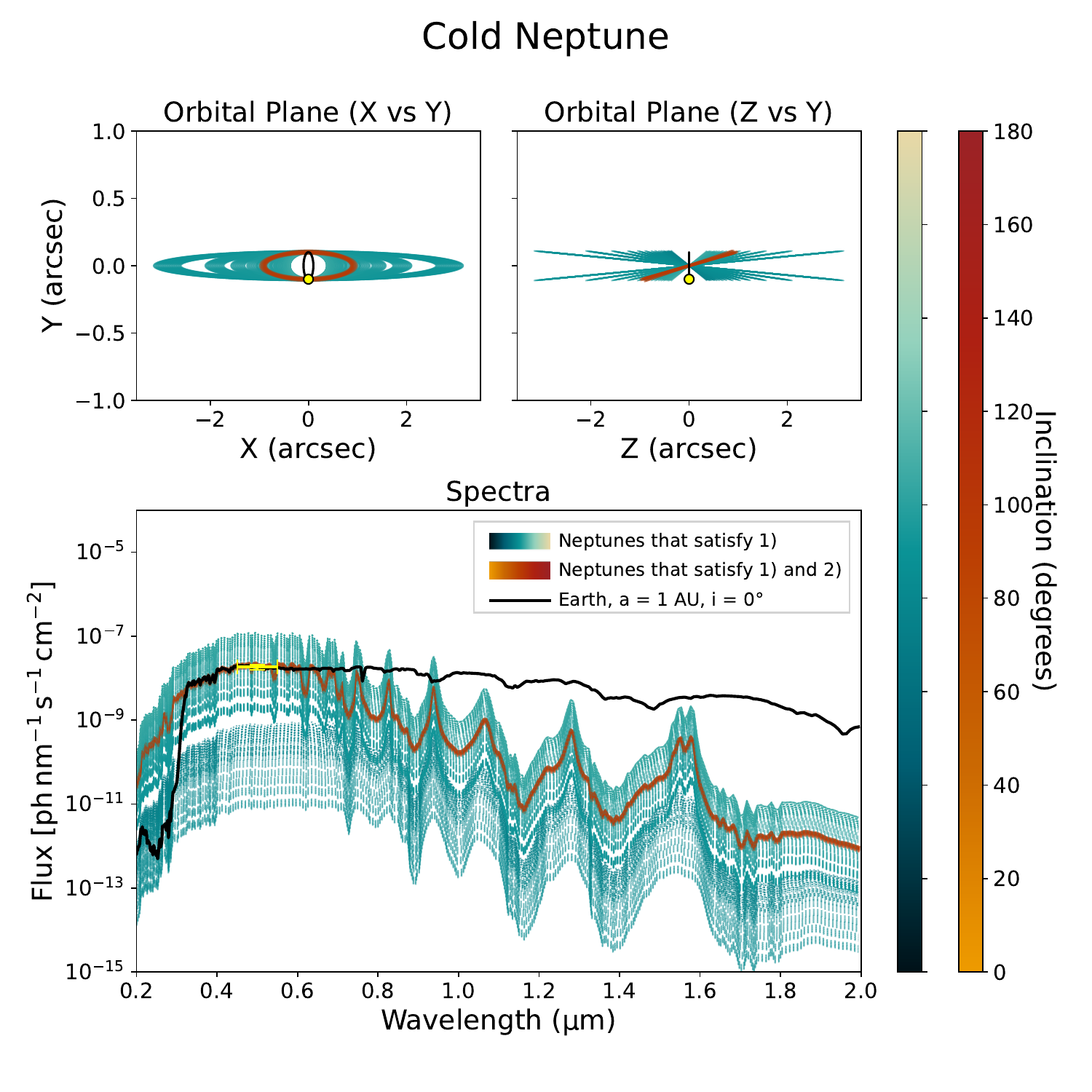}
    \caption{Orbits (\emph{top left panel}: $\hat{X}-\hat{Y}$ orbital projection; \emph{top right panel}: $\hat{Z}-\hat{Y}$ orbital projection) and fluxes (\emph{bottom panel}) of the cold Neptunes that are compatible with the assumed detection point, compared to the orbit and flux of an Earth at quadrature. In black, the orbit (top panels) and the flux (bottom panel) of the Earth on a face-on orbit. In yellow-green hues, the orbits (top panels) and fluxes (bottom panels) of all cold Neptunes that are compatible with the detected position (condition 1), color-coded according to the inclination of their orbits.  In red hues, the orbits (top panels) and fluxes (bottom panel) of the cold Neptunes that also yield a comparable flux at 500 nm (condition 2). In the top panels, the projected distance is shown as a yellow scatter plot. In the bottom panel, the photometric spectrum of the Earth template calculated at 0.5 $\mu$m with a 20\% bandwidth (horizontal errorbar), and assuming \snr=7  (vertical errorbar) is shown as a yellow errorbar point. }
    \label{fig:coldneptunesorbitsspectra}
\end{figure*}

In Figures \ref{fig:warmneptuneorbitsspectra} and \ref{fig:coldneptunesorbitsspectra} we show the orbits and the fluxes of the selected warm and cold Neptunes respectively, compared to the reference Earth. The planets that satisfy the orbital condition  are shown in  green-yellow hues, while the planets that also satisfy the flux condition upon comparison with the Earth's photometric flux at 500 nm are shown in orange-red hues. For both samples, the color bars vary based on the orbital inclination. In the top panels, we display the $\hat{X}-\hat{Y}$ and $\hat{Z}-\hat{Y}$ projections of the selected orbits compared to the one of the reference Earth at quadrature. In the bottom panels, we plot the fluxes of the warm and cold Neptunes at their appropriate phase and semi-major axis.

In Figure \ref{fig:warmneptuneorbitsspectra}, we can observe once more that the majority of the plausible orbits lie in configurations that are very close to that of the Earth in quadrature ($a\approx 1 $ AU, $i\approx~0-180^\circ$).  We observe that, for models at inclinations close to $0^\circ$ and $180^\circ$ (i. e., light yellow and the dark green in the color bar), the fluxes are entirely overlapping, demonstrating that the two populations of planets are actually in a similar phase configuration.  We also note that the fluxes of this subset of planets is the highest of the whole sample, since the smaller semi-major axis and a phase close to quadrature ($\phi=90^\circ$) result in the most stellar radiation scattered to the observer by the planet.  

Of all the simulated warm Neptunes (1167 models), 18 have a compatible flux at 500 nm with that of the Earth at quadrature ($1.5\%$ of the model sample). These models have a semi-major axis equal to 2.25 AU, inclination $\bar{i}\approx 65^\circ$, and true anomaly $t\approx 180^\circ$, resulting in a crescent phase $\phi\approx155^\circ$. These planets are therefore mostly dark, with only a small fraction of illuminated planetary atmosphere producing comparable flux to that of the Earth. The spectra of the warm Neptunes closest matching in flux (red hues) show higher NUV fluxes compared to the Earth, as well as deeper methane lines in the NIR. However, the continuum is comparable to that of the Earth at VIS/NIR wavelengths. The warm Neptune that best fits the Earth flux at 500 nm is the model at $a=2.25\ AU$, $\bar{i}=65^\circ$, $t=182^\circ$, with phase $\phi=155^\circ$ (1.2\% difference between $F^{Warm\ Neptune}_{500,20\%}$ and $F^{Earth}_{500,20\%}$).

Concerning the cold Neptune population (Figure \ref{fig:coldneptunesorbitsspectra}), we notice that there is no substantial variation in inclination, with all orbits lying at inclinations close to edge-on ($\bar{i}=[70^\circ,110^\circ]$). For this reason, variations in color according to the color bar are less noticeable. We observe a large variation in flux in VIS wavelengths: the planets at true anomaly close to $0^\circ-360^\circ$ (at negative values of $\hat Z $) have most of their disk illuminated and reflect the most stellar flux, while planets at true anomaly close to $180^\circ$ (at positive values of $\hat{Z}$) are almost in a ``nightside-only'' configuration, with little to no flux reflected. 

Of all the 268 cold Neptune models, 7 have fluxes at 500 nm that are similar to that of the Earth ($2.6\%$ of the model sample). These Neptunes have a semi-major axis between 8.75 and 9.50 AU, an inclination  $\bar{i}\approx 96^\circ$, and a true anomaly of $0^\circ$, which then corresponds to phases from $\phi\approx7^\circ$ (gibbous phase). These planets are far from the star and mostly illuminated, achieving comparable fluxes to that of the Earth at quadrature in the VIS range. However, stark differences in flux appear in the NIR regime: Neptune-like atmospheres become dark in this region due to the presence of strong CH$_4$ lines. Differences are also noticeable in the NUV, due to the lack of ozone. The cold Neptune that best fits the Earth flux at 500 nm is the model at $a=9.00\ AU$, $\bar{i}=97^\circ$, $t=0^\circ$, with resulting phase $\phi=7^\circ$ (2.5\% difference between $F^{Cold\ Neptune}_{500,20\%}$ and $F^{Earth}_{500,20\%}$).

\section{Noise Modeling}\label{sec:noise}

\begin{table*}
    \centering
    \begin{tabular}{p{6cm}p{2.2cm}p{8.8cm}}
       \hline Parameter  & Value & Meaning\\\hline\hline
       \multicolumn{3}{c}{Observation}\\\hline
        \texttt{wavelength} [$\mu$m] & 0.500 & Wavelength of the discovery observation ($\lambda_{discovery}$)\\
        \texttt{snr} & 7 & Desired \snr~at $\lambda_{discovery}$\\
        \texttt{CRb\_multiplier} & 2 & Factor $\alpha$ for PSF subtraction method (2 for Angular Differential Imaging) \\
        \texttt{distance} [pc] & 10 & Distance $d$ of the system \\
        \texttt{stellar\_radius} [$R_\odot$] & $1$  & Stellar radius\\
        \texttt{nzodis} [zodi] & 3 & Exozodi multiplier \\
        \texttt{ra} & 236.01$^\circ$ & Right ascension\\
        \texttt{dec} & 2.52$^\circ$ & Declination\\
        \texttt{angular\_separation} [arcsec] & 0.1& Separation (at $d$) \\\hline
       \multicolumn{3}{c}{Telescope}\\\hline
       \texttt{diameter} [m] & 7.870  & Circumscribed diameter $D_{circ}$ of telescope aperture \\
       \texttt{unobscured\_area} [\%] & 87.900 & Percentage of unobscured collecting area \\
       \texttt{photometric\_aperture\_radius} [$\lambda/D_{circ}$] & 0.850 & Photometric aperture radius  \\
              \texttt{toverhead\_multi} & 1.100& Multiplicative overhead time $\tau_\mathrm{multi}$  (refining the dark hole)\\
       \texttt{toverhead\_fixed}  [s] & 8250  & Static overhead time $\tau_\mathrm{static}$ (slew, settling, digging the dark hole)\\
       \texttt{telescope\_optical\_throughput} & 0.823 & Throughput of the telescope optics\\
        \texttt{temperature} [K] & 290 & Temperature of the optics\\
        \texttt{T\_contamination} &  0.950& Effective throughput due to contamination \\\hline
        
        \multicolumn{3}{c}{Coronagraph}\\\hline
\texttt{Istar} & $1.700\cdot10^{-15}$& Stellar intensity scalar ($I_{star}$, on-axis PSF), product of the noise floor contrast of the coronagraph (uniform over dark hole) and the peak value of the off-axis PSF. \\
        \texttt{noisefloor} &$5.119 \cdot 10^{-17}$ & Noise floor ($NF$) of the coronagraph \\
        \texttt{photometric\_aperture\_throughput} & 0.297 & Fraction of light entering the coronagraph that ends up within the photometric core of the off-axis (planet) PSF assuming perfectly reflecting/transmitting optics\\
        \texttt{skytrans} & 0.650 & Coronagraph’s performance when observing an infinitely extended source \\

        \texttt{coronagraph\_optical\_throughput} & 0.449 & Throughput of all coronagraphic optics\\\hline
        
         \multicolumn{3}{c}{Detector}\\\hline
\texttt{DC} [$e^-\ pix^{-1} s^{-1}$] & $3.000\cdot10^{-5}$ & Dark current\\
         \texttt{RN} [$e^-\ pix^{-1} read^{-1}$] & 0  & Read Noise\\
         \texttt{tread} [s] & 1000  & Time between reads\\
        \texttt{CIC} [$e^-\ pix^{-1} ph^{-1}$] & $1.300\cdot10^{-3}$  & Clock-induced charge\\
         \texttt{QE} & 0.900 & Quantum efficiency\\
         \texttt{dQE} & 0.750 & Effective QE due to degradation\\
        \hline
    \end{tabular}
    \caption{{Parameters used in the Exposure Time calculations (assuming the ``Toy Model'' observatory preset) and their keywords within \pyedith. }For parameters that are not dimensionless, units are expressed. }
    \label{tab:etcparameters}
\end{table*}

Having identified the cold and warm Neptunes that could reasonably be in an orbital configuration consistent with the detected planet (condition 1, Section \ref{sec:orbits}), and whose flux at 500 nm is similar to that of the Earth at quadrature (condition 2, Section \ref{sec:fluxes}), we now investigate whether additional photometric information can distinguish between these scenarios. This requires modeling the expected noise over the full wavelength range.

For this work, we developed a new exposure time calculator (ETC) called \pyedith~ (Python-based Exposure Direct Imaging Timer for HWO). This is a coronagraph ETC made specifically for the Habitable Worlds Observatory. \pyedith~ is designed to handle coronagraphic observations in photometry or spectroscopy mode and it enables simulations of both user-defined observatory concepts as well as the Exploratory Analytic Cases (EACs) considered in the architecture trades. 

\pyedith~ was originally adapted from the ETC within the Altruistic Yield Optimization code \citep[AYO,][]{2014ApJ...795..122S}, which is currently being used for yield estimates in the HWO community. The ETC module of AYO, written in IDL and C, has been released open-source\footnote{\url{https://starkspace.com}}. As part of our \pyedith~ efforts, we first translated the code into Python, then we implemented the seamless ingestion of both the EAC YAML files\footnote{\url{https://github.com/HWO-GOMAP-Working-Groups/Sci-Eng-Interface/tree/main/hwo_sci_eng}} and the coronagraph Yield Input Packages (YIP) files that are produced by different teams developing coronagraph designs and technologies for HWO. We describe in detail how to use \pyedith~and the noise terms that are considered for the exposure time calculation in \citet{alei_2025_17917472,pyedith2025} and in the online documentation\footnote{\url{https://pyedith.readthedocs.io/en/latest/index.html}}.

A specific feature of \pyedith~was developed specifically with this study in mind: in photometry mode, \pyedith~can calculate the exposure time $\tau$ necessary to reach the desired signal-to-noise ratio in a primary bandpass (in this case the discovery bandpass $\lambda_{discovery}=0.5\ \mu m$), and then use that exposure time to determine the corresponding \snr~for potential parallel bandpasses. This way, we can simulate simultaneous photometric observations in multiple bandpasses in a self-consistent manner. 

In Table \ref{tab:etcparameters} we list the parameters used in the noise simulation; these are the same for all simulated scenarios. In our simulations, we used a simplified  observatory through the special ``Toy Model'' class in \pyedith. The ``Toy Model'' observatory is a simplified version of the first Exploratory Analysis Case (EAC1) considered for HWO. The mirror dimensions follow the prescribed ones for the EAC1 case, but all throughputs and coronagraph specifications are averaged over wavelength, so that every value at every wavelength is constant. This is an approximation, but it is a valuable one, since it allows us to provide the community with a more generalized answer that would inform future studies on coronagraph and detector specifications, which we will also explore in a future study (see Section \ref{sec:limitations}). A Toy Model observatory was also used in the validation of other existing ETCs for HWO in \citet{2025arXiv250218556S}.

We consider the Earth at quadrature and the best warm and cold Neptunes identified in Section \ref{sec:fluxes}. We first calculate the exposure time $\tau_{discovery}$ that would correspond to \snr=7 at the discovery bandpass (centered at $\lambda_{discovery}=0.5\ \mu m$ and with a 20\% bandwidth) for the three planets in our study. This value is $\approx 3.17\ hr$. As a reminder, all planets have comparable $\tau_{discovery}$ by construction, as we purposely chose the Neptune planets whose flux in the $\lambda_{discovery}$ bandpass was the closest to the Earth's flux (see Section \ref{sec:fluxes}). 

Then, for each planet, we step across the full wavelength range from the UV to the NIR, calculating the \snr~that would result from observing in a bandpass centered at different wavelength positions and for 5\%, 10\% and 20\% bandwidth for the same exposure time $\tau_{discovery}$. This way, we can simulate what would be the quality of the observation on a parallel bandpass if we allocate exactly the same exposure time necessary to achieve the target \snr~in the detection band. For each bandpass, we calculate the average planetary and stellar fluxes that would be acquired in a photometric observation over the whole bandpass range and use those for the noise calculation within the ETC.

\pyedith~provides us with an uncertainty associated with the photometric observation for each wavelength data point. We can then identify the bandpasses that would produce the largest differentiation between the Earth and either of the two Neptune-like planets by evaluating the difference between the estimated flux for the Earth and the  Neptune, and dividing it by their uncertainty. We refer to this metric, the statistical significance of the difference in flux between the Earth and each Neptune case, as $S$:

$$S= \frac{|F_{Earth}-F_{Neptune}|}{\sigma}\mathrm{, where}\  \sigma=\sqrt{\sigma_{Earth}^2+\sigma_{Neptune}^2}\label{eq:significance}$$

This approach has been widely used in literature to determine the detectability of a spectral feature, or as a way to discriminate between two models \citep[see e.g.,][]{LustigYaeger2019, Bixel2021, 2024arXiv240108492A}. We would consider the two models qualitatively distinguishable in a specific bandpass if $S\ge3$ for that bandpass, and reliably distinguishable if $S\ge5$. This means that the difference between the two models is greater than three or five times the uncertainty.

\subsection{Results}

\begin{figure*}[htbp]
    \centering
    \includegraphics[width=\linewidth]{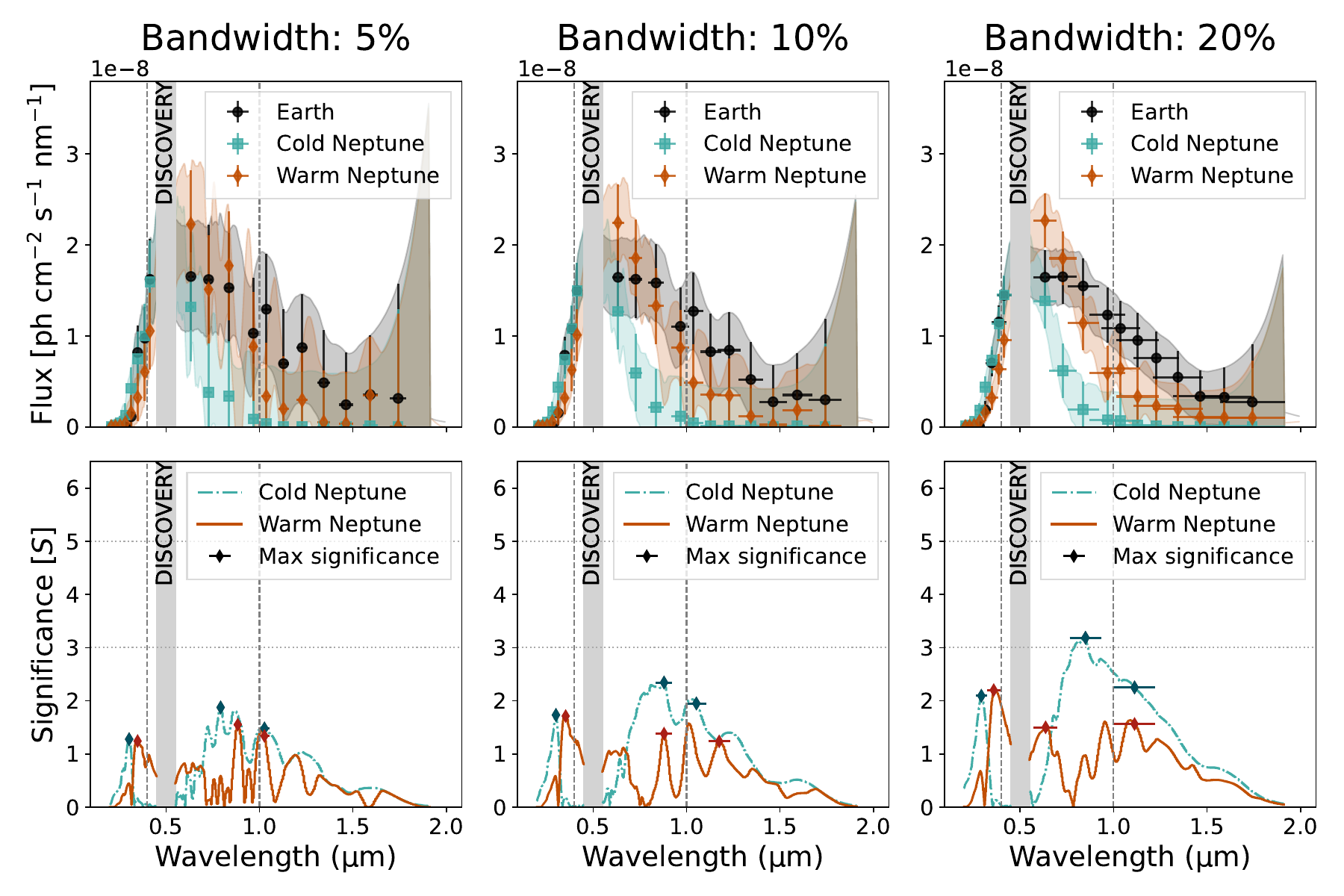}
    \caption{\emph{Top}: High-fidelity noise simulations for potential additional bandpasses that could be observed at the same time as the discovery observation for different assumed bandwidths. The shaded areas represent the uncertainty calculated from the ETC for the entire bandpass sweep, of which only a few points are shown. The horizontal errorbar indicates the width of the bandpass. The models are color-coded as follows: Earth in black, cold Neptune in aqua, warm Neptune in orange. The discovery bandpass is shown as a gray shaded area, and two gray dotted lines separate the UV, VIS, and NIR channels. \emph{Bottom}: Significance $S$ of the entire bandpass sweep. As diamond points, the central wavelengths that would yield the maximum $S$ value within the channel, provided they do not intersect with the channel edges and the discovery bandpass. The horizontal errorbar indicates the width of the bandpass. The models are color-coded as follows: difference between Earth and cold Neptune in aqua, difference between Earth and warm Neptune in orange. The discovery bandpass is shown as a gray shaded area; two dashed lines separate the UV, VIS, and NIR channels; two dotted lines illustrate the $S=3$ and $S=5$ differentiation thresholds.
}
    \label{fig:noisemodelingsignificance}
\end{figure*}

  In Figure \ref{fig:noisemodelingsignificance} we plot the simulated photometric measurements for the Earth at quadrature, the best warm Neptune, and the best cold Neptune, as well as the $S$ value for each bandpass position, for the varying bandwidths (5\%, 10\%, 20\%).

We find that the photometric flux measurements at smaller bandwidths are more sensitive to morphology in the planetary spectra, and $S$ is lower compared to the largest bandwidth explored (20\%). Increasing the bandwidth means reducing the sensitivity over specific line shapes in favor of acquiring more photons, increasing the derived \snr~and therefore reducing the uncertainty on each measurement and improving $S$. This can be confirmed also by qualitatively examining the selected bandpass results that we show on the top row of Figure \ref{fig:noisemodelingsignificance}: with increasing bandwidth, we are more and more able to discriminate between the various models. 

This is especially visible in the cold Neptune scenario, which nearly doubles in significance at its peak around 850 nm when quadrupling the bandwidth (5\% to 20\%, from $S\sim1.8$ to $\sim3.2$ at the peak). At wavelengths between 790 and 850 nm at 20\% bandwidth, the cold Neptune can be more easily differentiated from the Earth compared to the warm Neptune case, with $S>3$. The strong significance at this wavelength range reflects the methane and cloud absorption features that distinguish Neptune from the Earth.  Observing with one or more secondary bandpass(es) in this range would effectively differentiate the two models (see Section \ref{sec:observations}).

The warm Neptune is overall harder to differentiate from the Earth, with $S\sim1.5$ at best independent of the bandwidth. However, for the 20\% bandwidth case more wavelength regions across the whole range achieve higher $S$ values, as smaller peaks in significance join to form a larger one. The limited discrimination potential stems from similarities in the two spectra (see Section \ref{sec:fluxes}).  Longer observing times would increase the \snr~and might allow for an easier differentiation, which otherwise is more challenging compared to the cold Neptune case (see Section \ref{sec:discussion-snr}).

For both models, there is a diminishing return from observing at NIR wavelengths $\gtrsim 1.2\ \mu m$ for which the noise becomes larger than the fluxes to be measured.

\section{Optimizing Parallel Channel Positions}\label{sec:observations}

After establishing the potential bandpasses that might enhance the differentiation between the Earth and the Neptunes, we now investigate what combinations of multiple (2 or 3) simultaneous photometric bandpass measurements could break these degeneracies.

Based on our current knowledge of the potential coronagraph channels that HWO might have, we assume two hard wavelength cutoffs to separate the channels: $0.4\ \mu m$ separates the UV and the VIS channels, and $1.0\ \mu m$ is the threshold between the VIS and NIR channels. These wavelength boundaries reflect current assumptions for HWO's coronagraph detector architectures and delineate different parallel coronagraph channels that would be handled by separate optical systems and detectors. Given these constraints, we can imagine a few potential options for multi-bandpass, simultaneous photometric observations. 

\begin{figure}
    \centering
    \includegraphics[width=\linewidth]{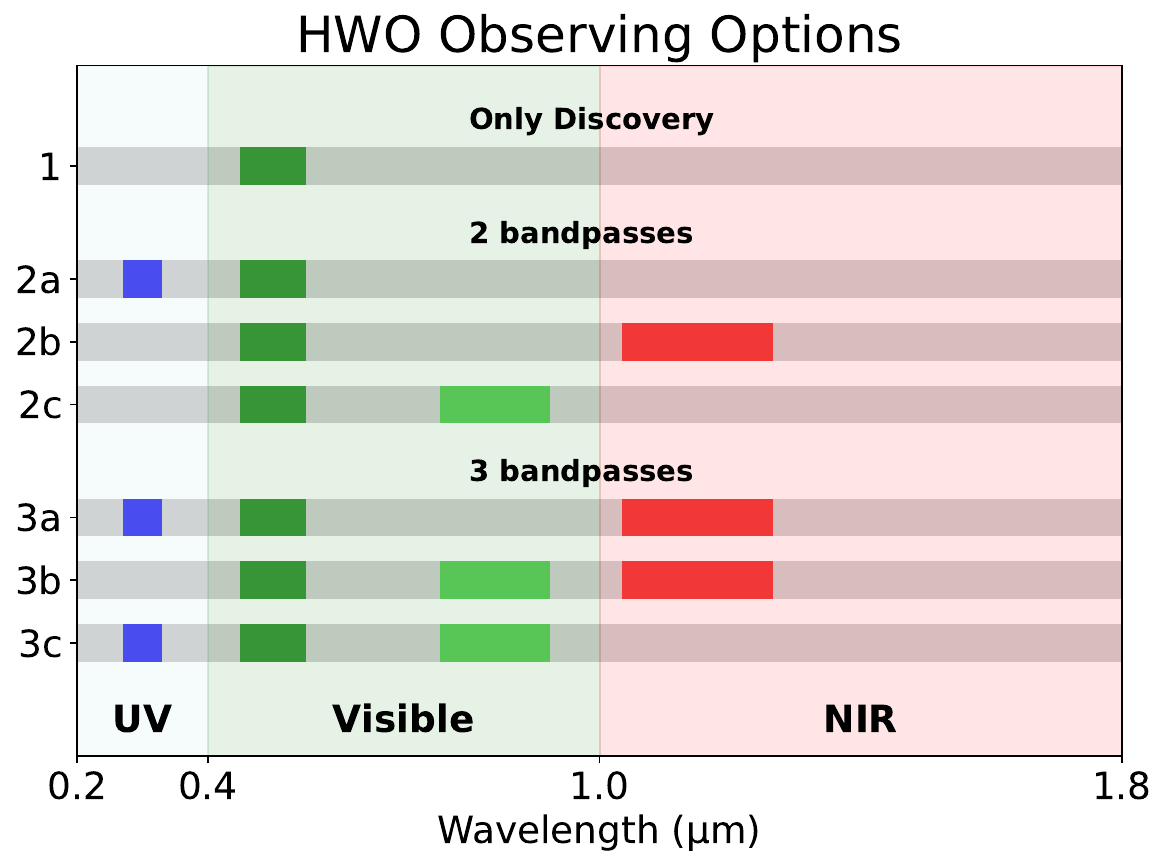}
    \caption{Potential observing strategies for HWO during the first visit. Channels and bandpasses are color-coded: UV in blue, VIS in green, and NIR in red.}
    \label{fig:options}
\end{figure}

In Figure \ref{fig:options} we identify the limited set of possible combinations of parallel channels that could be available with increasing technical complexity. Option 1 considers just one bandpass in the VIS range (the discovery bandpass); this option will not be sufficient in this study by definition, since we assume that all fluxes are essentially equivalent in the discovery bandpass. Option 2 considers two parallel bandpasses and has three different flavors: 2a, where the two bandpasses are in the UV and VIS range; 2b, with one bandpass in the VIS and one in the NIR range; 2c, with two bandpasses in the VIS range. Option 3 considers three simultaneous bandpasses and also has three flavors: 3a, with a bandpass in each channel; 3b, with two bandpasses in the VIS and one in the NIR range; 3c with two bandpasses in the VIS and one in the UV range.

We identify the best parallel bandpass position for each scenario (Earth versus Warm Neptune, Earth versus Cold Neptune) as the ones that maximize $S$ for each channel (UV, VIS, NIR) and for every simulated bandwidth (5\%, 10\%, 20\%). To do this, we require that the photometric bin is fully included within a single channel and does not overlap with the discovery bandpass. This choice is motivated by technology considerations: 1) two bandpasses cannot overlap during the same observation, and 2) different channels would feed into different coronagraphs and detectors, so there should be a hard cutoff that separates each channel from the others. We assume idealized filters with perfect transmission between the edges of the bandpass and no transmission outside of that range, and that we can freely select the central wavelength. This will surely not be the case in reality, where specific detector responses must be accounted for, and a discrete set of filters will eventually be selected. We discuss the implications of these assumptions in Section \ref{sec:limitations}.

When comparing the different options, we calculate the total significance of the difference in flux measurements for each option by calculating the quadratic sum of the $S$ values for all of the bandpasses ($S_{bandpass}$) as described by the different options (hereafter $S_{option}$):

$$S_{option}= \sqrt{\sum_{option}S_{bandpass}^2} $$

 The higher the value of $S_{option}$ is, the better that option can distinguish between the Earth and a Neptune. This metric will be useful in distinguishing the preferable ``flavor'' of each option and the option(s) most promising for a general differentiation of the model scenarios.

In the case where three simultaneous bandpasses are explored, we can also compare the difference in the photometric flux observed between two different pairs of bandpasses. For every flavor of Option 3, we can select the three bandpasses ($1,\ 2,\ 3$) and calculate the differences in the fluxes that would be acquired in each bandpass, each with its own uncertainty. We can then compare $(F_1/F_2)$ versus $(F_3/F_2)$, as these values represent the difference in color between pairs of bandpass. This is done in a similar fashion in previous literature studies \citep[see e.g.][]{JKT2016,2018AJ....156..158B,Smith2020}. The three bandpasses are not necessarily ordered by increasing wavelength, but we compare each secondary bandpass with respect to the discovery bandpass: $F_2$ is centered at 500 nm, while the $F_1$ and $F_3$ are centered on the secondary bandpass, regardless of where they are located.

In this parameter space we can then calculate the distance between two models and evaluate how effective each 3-bandpass ``flavor'' is in differentiating each planet. We do so by calculating the vector distance between pairs of models in color-color space (where we denote $y=F_1/F_2$ and $x=F_3/F_2$ and we label the Earth as $E$ and any Neptune as $N$):

\begin{equation}
    d = \sqrt{(x_{E}-x_{N})^2+(y_{E}-y_{N})^2}
\end{equation}

And its related uncertainty:
\begin{equation}
   \Delta d = \sqrt{\sum_{z}\left(\frac{\partial d}{\partial z}\right)^2\sigma_{z}^2}\mathrm{,\ where}\ z=[x_E, x_N, y_E, y_N]
\end{equation}
    
We use the ratio of the distance over its uncertainty as a metric to distinguish the Earth from any other Neptune:
\begin{equation}
    r=\frac{d}{\Delta d}
\end{equation}

We determine a planet to be qualitatively distinguished from the Earth if $r\ge3$ and decisively distinguished if $r\ge5$, meaning that the separation in color-color space is more than three or five times greater than the uncertainty on the measurement.

\subsection{Results}

    \begin{table}[htbp]
    \centering
    \begin{tabular}{cccc}
    \hline
    \multicolumn{4}{c}{{Warm Neptune}}\\
    \hline
    Channel & Bandwidth & $\lambda_{bandpass}$ ($\mu$m) & $S_{bandpass}$ \\
    \hline\multirow{3}{*}{UV (0.2-0.4 $\mu$m)} & 5\% & 0.35 & 1.24 \\
& 10\% & 0.35 & 1.71 \\
& 20\% & 0.36 & 2.20 \\
\hline
\multirow{3}{*}{VIS (0.4-1.0 $\mu$m)} & 5\% & 0.88 & 1.55 \\
& 10\% & 0.88 & 1.38 \\
& 20\% & 0.64 & 1.50 \\
\hline
\multirow{3}{*}{NIR (1.0-1.8 $\mu$m)} & 5\% & 1.03 & 1.34 \\
& 10\% & 1.17 & 1.24 \\
& 20\% & 1.11 & 1.56 \\
\hline
Discovery & 20\% & 0.5 & 0.07\\\hline\hline
    \multicolumn{4}{c}{{Cold Neptune}}\\
    
    \hline
    Channel & Bandwidth & $\lambda_{bandpass}$ ($\mu$m) & $S_{bandpass}$ \\
    \hline\multirow{3}{*}{UV (0.2-0.4 $\mu$m)} & 5\% & 0.30 & 1.28 \\
& 10\% & 0.30 & 1.73 \\
& 20\% & 0.29 & 2.09 \\
\hline
\multirow{3}{*}{VIS (0.4-1.0 $\mu$m)} & 5\% & 0.79 & 1.88 \\
& 10\% & 0.88 & 2.34 \\
& 20\% & 0.85 & 3.19 \\
\hline
\multirow{3}{*}{NIR (1.0-1.8 $\mu$m)} & 5\% & 1.03 & 1.48 \\
& 10\% & 1.05 & 1.95 \\
& 20\% & 1.11 & 2.25 \\
\hline
Discovery & 20\% & 0.5 & 0.12\\\hline\hline
    \end{tabular}
    \caption{{Bandpasses whose significance is maximized within each channel for the warm and cold Neptune scenarios.} Rows show the different channels and bandwidths that were modeled, as well as the calculated central wavelengths and the corresponding significances. }
    \label{tab:bandpassesneptunes}
    \end{table}

    \begin{figure*}
\includegraphics[width=\linewidth]{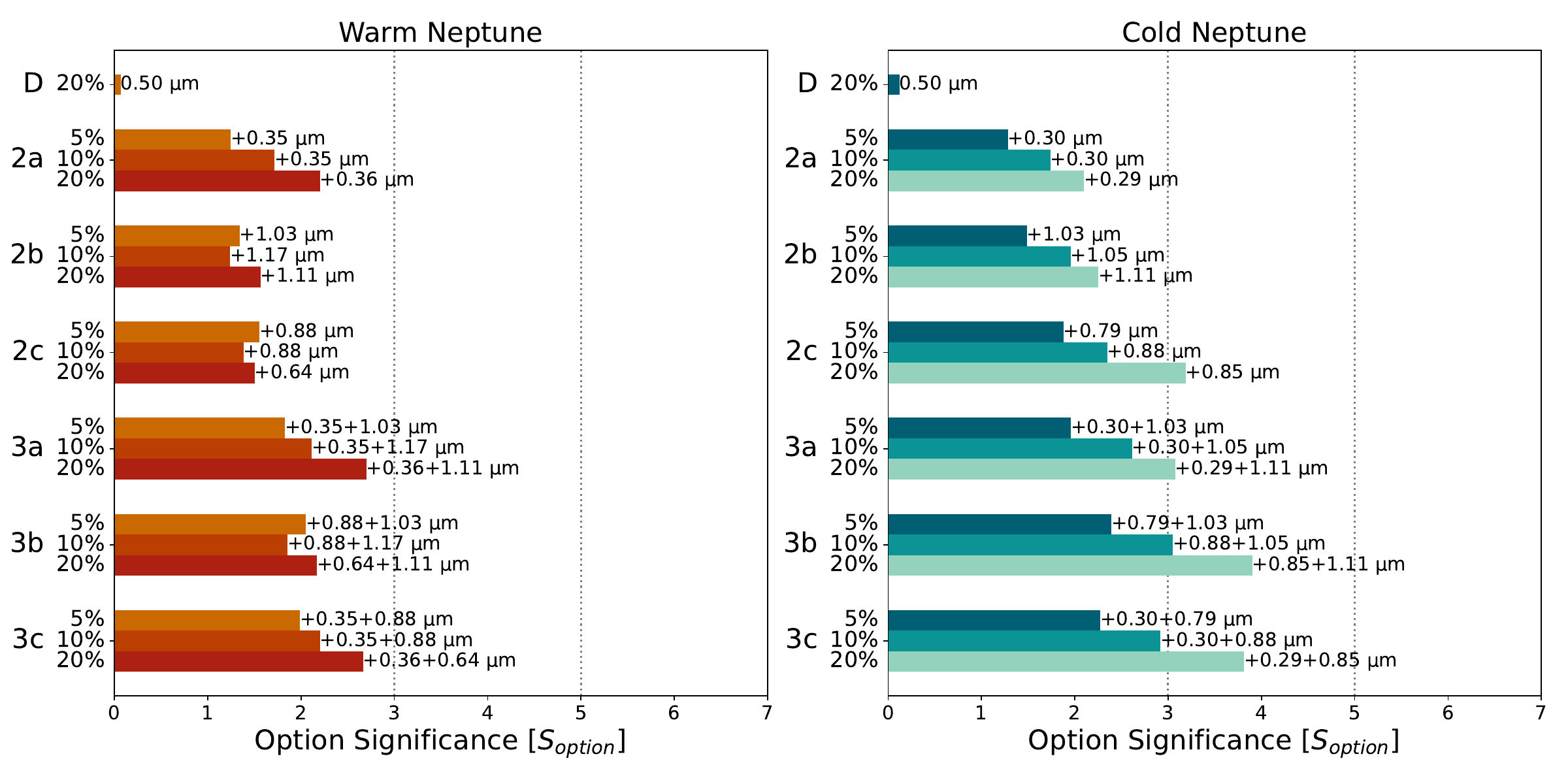}

    \caption{Option significance for the various flavors of Options 2 and 3 for the Warm Neptune  case (left) and the Cold Neptune case (right) at varying bandwidths, compared with the discovery-only case (D). For each option and bandpass, we annotated the additional bandpasses that would be needed to reach that significance. Two dotted lines illustrate the $S=3$ and $S=5$ differentiation thresholds. }
    \label{fig:soption}
\end{figure*}

In {Table \ref{tab:bandpassesneptunes} }we list the central bandpass wavelength $\lambda_{bandpass}$ that maximizes the significance in each channel for each bandwidth, as well as the value of the significance in that bandpass $S_{bandpass}$. These bandpasses are also shown in Figure \ref{fig:noisemodelingsignificance} as color-coded diamonds for each scenario, with errorbars labeling the width of the bandpass.

By comparing the selected bandpasses with the overall behavior of the significance (Figure \ref{fig:noisemodelingsignificance}), we notice that there are specific local maxima which often coincide with the best bandpass selected for that channel. For the warm and the cold Neptune, UV peaks at 350 and 300 nm respectively for all bandpasses; in the VIS range, wavelengths around 800-880 nm are preferred; in the near-infrared, the significance $S$ peaks around 1-1.2 $\mathrm{\mu m}$. There is a noticeable exception for the warm Neptune scenario: a strong peak in $S$ around 1.03 micron cannot be leveraged by the 10\%- and 20\%-wide bandpasses, as the width of the bandpass would intersect the hard threshold wavelength that is defined to separate the VIS and NIR channels. A different cutoff between VIS and NIR could provide enhanced differentiability between the Earth and warm Neptune (see Section \ref{sec:discussion-comparison}). Critically, the inability to leverage the higher $S$ values at longer VIS wavelengths constrains the VIS channel bandpass optimization for the Warm Neptune scenario to $\sim 640$ nm, which has lower significance and also is not in agreement with the corresponding VIS bandpass for the cold Neptune case. Around 640 nm (i.e., the preferred wavelength for the warm Neptune case at 20\% bandwidth) the cold Neptune scenario shows the lowest significance of all the VIS channel bandpass positions, so it is hard to differentiate the cold Neptune with respect to the Earth; on the other hand, it would be much easier to differentiate cold Neptune and Earth at 850-880 nm (highest $S$) but we are faced with a local minimum in $S$ for the warm Neptune scenario. This shows that it might be difficult to find a single bandpass combination (or observation strategy) that is convenient to differentiate Earth-like atmospheres from both warm and cold Neptunes.

In Figure \ref{fig:soption} we show the significance of the various configurations of the two and three simultaneous bandpass scenarios (Options 2 and 3), for varying bandwidths. We notice a general increase in $S_{option}$ by considering two or three simultaneous bandpasses compared to the discovery bandpass only (D). This is by construction, since we selected the warm and cold Neptune models based on the smallest difference in flux in the discovery bandpass. The significance is very low in this case, but not zero, because of the uncertainty on the simulated measurement (condition 2, see Section \ref{sec:fluxes}). 

We observe a general behavior with 20\% bandwidths preferred compared to smaller ones. This is always true for the cold Neptune, but some exceptions occur in the 2b and 2c options for the warm Neptune, which marginally prefer 5\% bandwidth compared to larger ones. This is because the smallest bandwidth scenario can at least partially leverage the peak around $\sim 1.03\ \mu m$, which is not available to wider bandwidth cases. Options that feature three bandpasses with the smallest bandwidth generally show a higher significance compared to a similar 2-bandpass option with 20\% bandwidth.

We also observe that, for the same bandwidth, most Option 3 cases show an increase in their total $S_{option}$ value compared to two-bandpasses scenarios, except for some specific cases.  One such example is Option 2c  compared to 3a for a cold Neptune at 20\% bandwidth ($S_{2c,20\%}=3.19$ and $S_{3a,20\%}=3.08$): having just one additional bandpass in the VIS range could be more valuable than having two additional bandpasses (one in the UV and one in the NIR). Having an additional bandpass in the NIR together with the one in the VIS range would further enhance the differentiability (Option 3b, the best option overall for a cold Neptune, $S_{3b,20\%}=3.90$).

In the Warm Neptune case none of the options produce $S>3$, meaning that the difference in flux between the Warm Neptune and the Earth is less than three times the corresponding noise (as described in \ref{sec:noise}). In the two-bandpass scenarios, Option 2a is generally preferred to the others for $\ge10\%$  bandwidth ($S_{2a,10\%}=1.72$,$S_{2a,20\%}=2.20$). When three bandpasses are available, Options 3a (UV+D+NIR) and 3c (UV+D+VIS) at 20\% bandwidth are almost equally preferred, with $S\sim2.7$.

\begin{figure*}[htbp]
    \centering
        \begin{subfigure}[b]{\textwidth}
        \centering
        \includegraphics[width=0.45\textwidth]{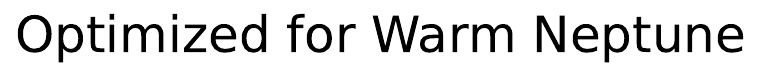}
    \end{subfigure}
    \begin{subfigure}[b]{\textwidth}
        \centering
        \includegraphics[width=0.93\textwidth]{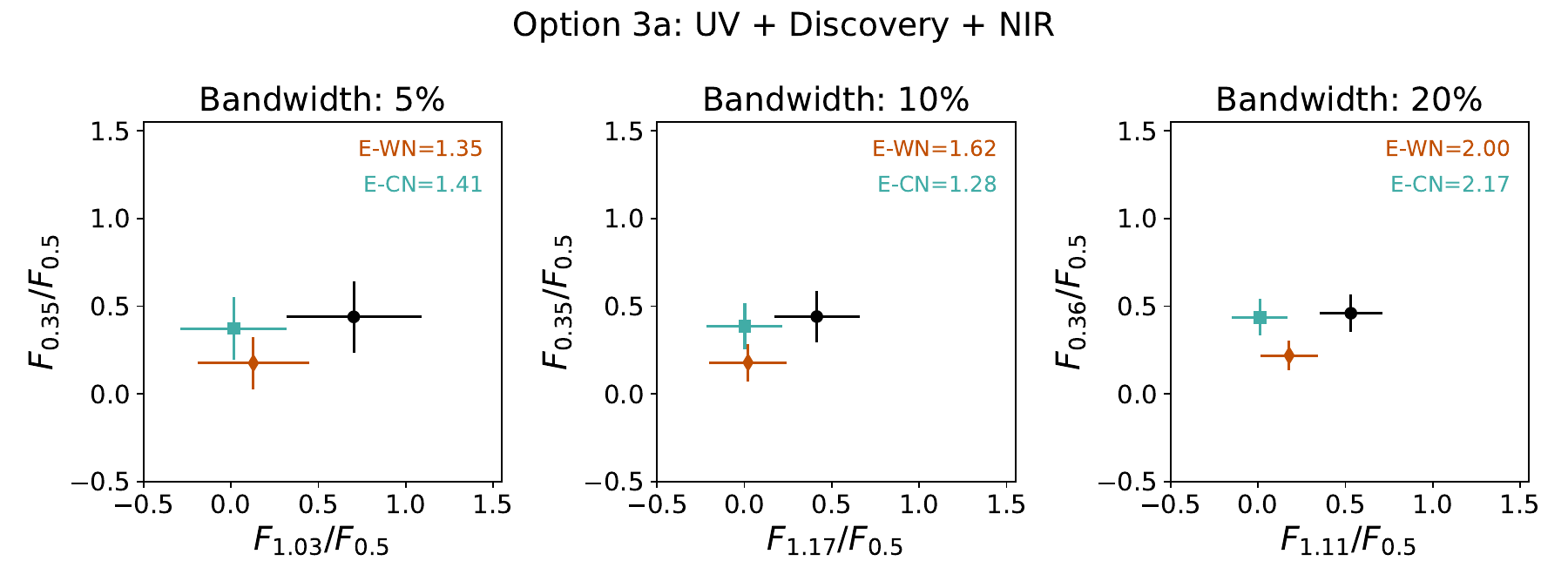}

    \end{subfigure}
    \hfill
    \begin{subfigure}[b]{\textwidth}
        \centering
        \includegraphics[width=0.93\textwidth]{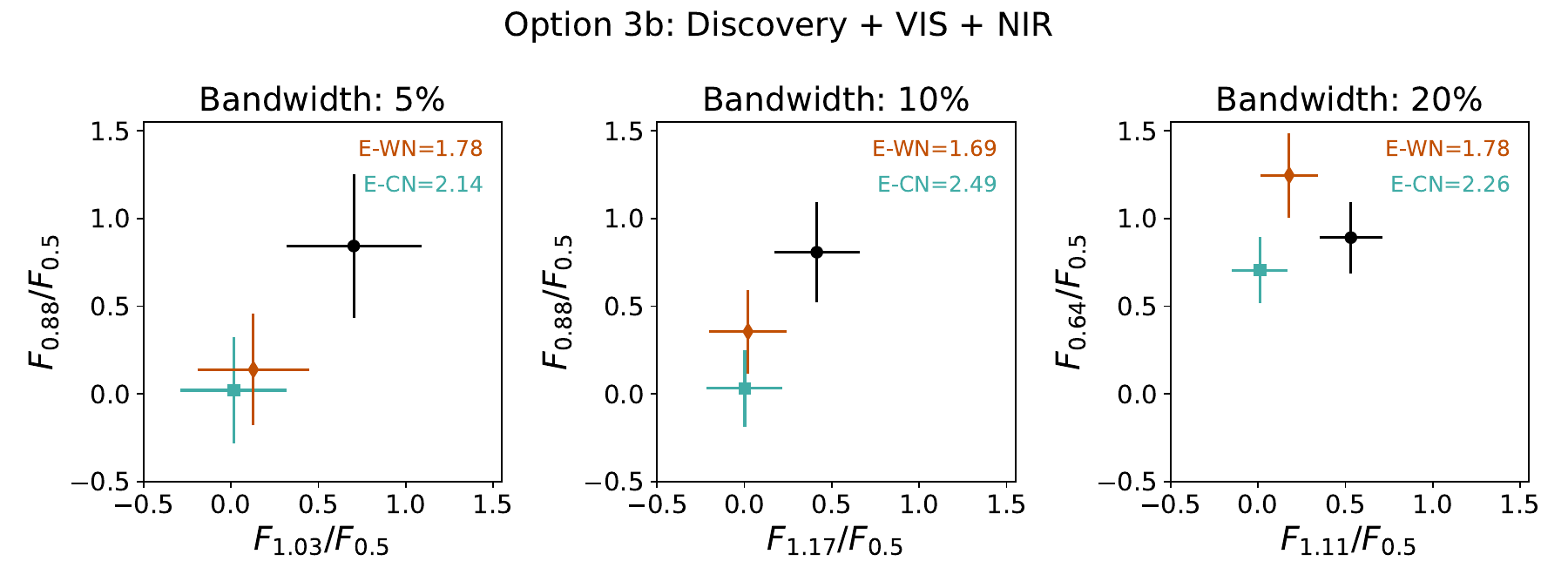}
    \end{subfigure}
    
    \vspace{1em}
    
    \begin{subfigure}[b]{\textwidth}
        \centering
        \includegraphics[width=0.93\textwidth]{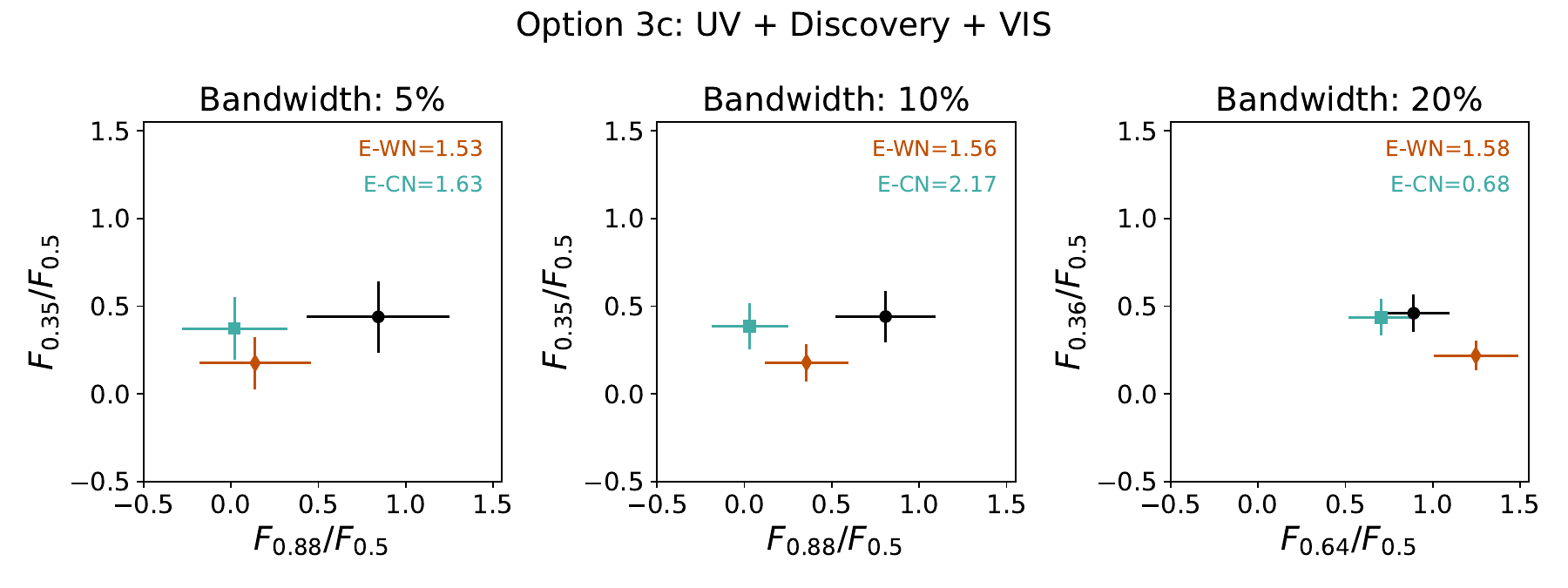}

    \end{subfigure}
    \begin{subfigure}[b]{\textwidth}
        \centering
        \includegraphics[width=0.5\textwidth]{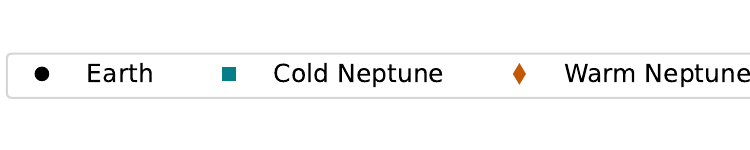}

    \end{subfigure}
    \caption{ Color-color diagrams for the 3-bandpass cases at varying bandwidth, optimized for warm Neptune differentiation. Points show Earth (black), cold Neptune (aqua), and warm Neptune (orange). At the top-right corner of each subplot, we show the ratio $r$ of the distance between two points relative to the corresponding uncertainty (E-WN: differentiation between Earth and warm Neptune, E-CN: differentiation between Earth and cold Neptune).
}
    \label{fig:colorcolor_warmneptune}
\end{figure*}

\begin{figure*}[htbp]
    \centering
    
        \begin{subfigure}[b]{\textwidth}
        \centering
        \includegraphics[width=0.45\textwidth]{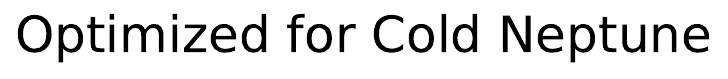}
    \end{subfigure}
    \begin{subfigure}[b]{\textwidth}
        \centering
        \includegraphics[width=0.93\textwidth]{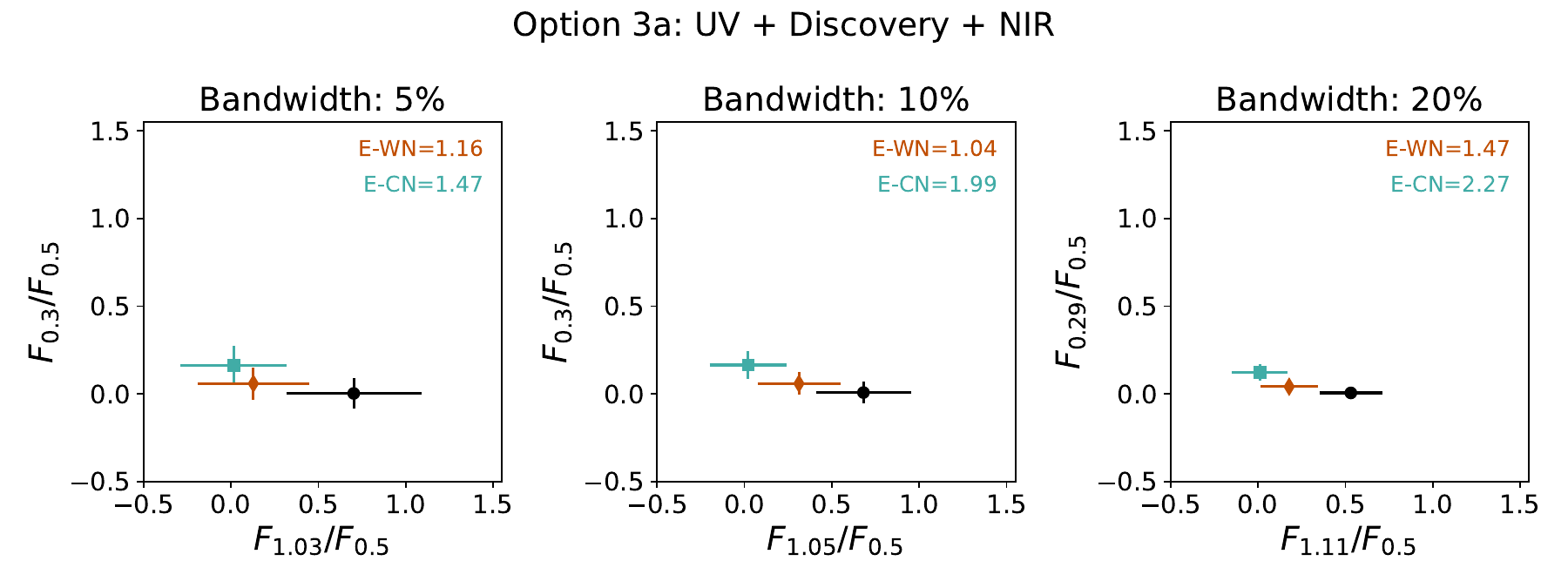}

    \end{subfigure}
    \hfill
    \begin{subfigure}[b]{\textwidth}
        \centering
        \includegraphics[width=0.93\textwidth]{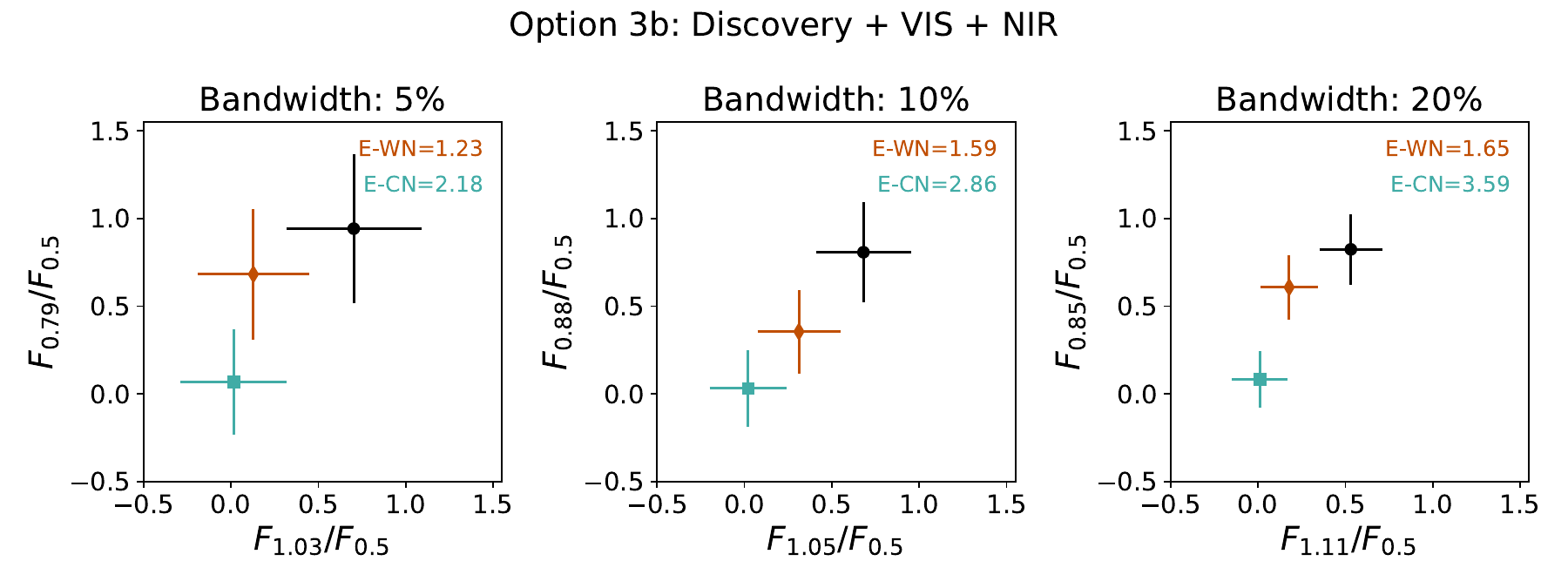}
    \end{subfigure}
    
    \vspace{1em}
    
    \begin{subfigure}[b]{\textwidth}
        \centering
        \includegraphics[width=0.93\textwidth]{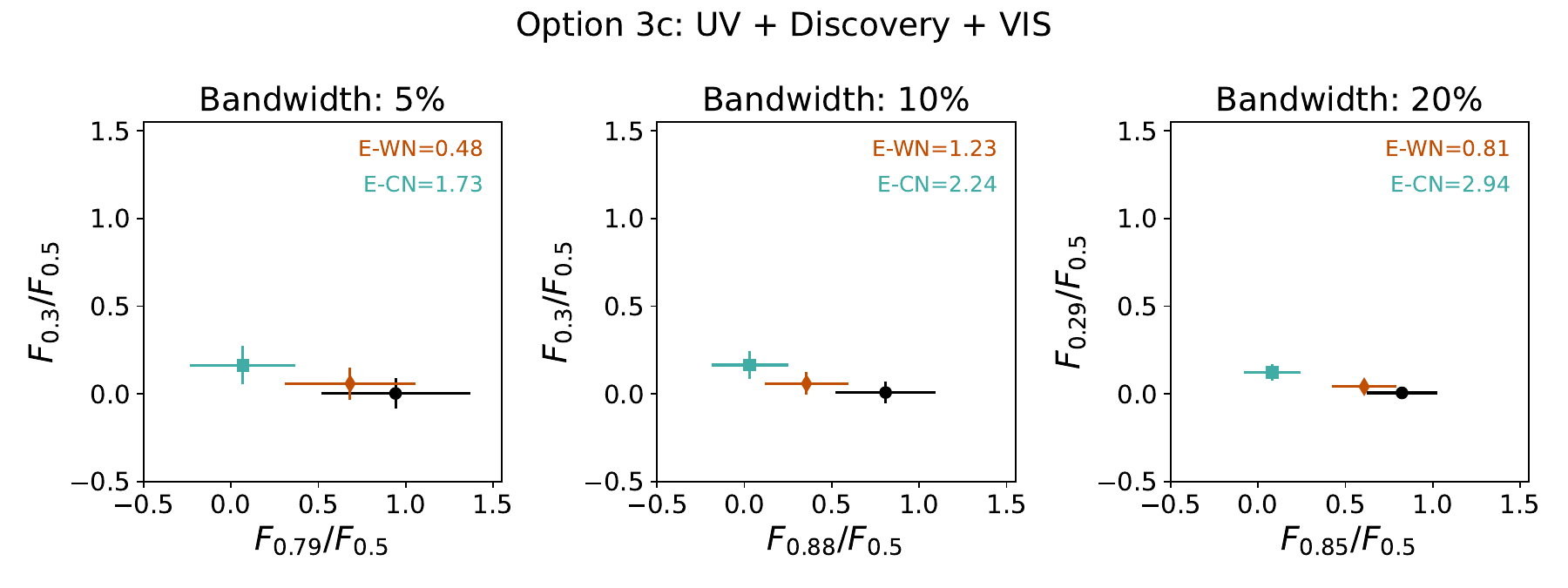}

    \end{subfigure}
    \begin{subfigure}[b]{\textwidth}
        \centering
        \includegraphics[width=0.5\textwidth]{colorcolor_legend.pdf}

    \end{subfigure}
    \caption{Color-color diagrams for the 3-bandpass cases at varying bandwidth, optimized for cold Neptune differentiation. Points show Earth (black), cold Neptune (aqua), and warm Neptune (orange). At the top-right corner of each subplot, we show the ratio $r$ of the distance between two points relative to corresponding uncertainty  (E-WN: differentiation between Earth and warm Neptune, E-CN: differentiation between Earth and cold Neptune).}
    \label{fig:colorcolor_coldneptune}
\end{figure*}

When considering three simultaneous observations, we can display our results in terms of color. In Figures \ref{fig:colorcolor_warmneptune} and \ref{fig:colorcolor_coldneptune} we show the color-color plots for the three flavors of Option 3, assuming the bandpasses {that maximize the differentiation significance of the warm Neptune case and the cold Neptune case respectively (see Table \ref{tab:bandpassesneptunes})}. In the top right corner of each subplot, we report the ratios $r$ between the Earth and the warm (cold) Neptune.  These plots show how a discrete choice of bandpasses (optimized for one model) can impact the detectability of the other. 

Disregarding for now the uncertainty associated with this measurement, we find that the planets are more effectively separated using narrow bandpasses: for the same option, the value of $d$ increases for decreasing bandwidths, resulting in points wider apart in color-color space. This has been observed also in \citet{JKT2016} for a similar study (see Section \ref{sec:discussion-comparison} for a more in depth comparison). However, the uncertainty on the flux measurements also becomes increasingly larger at smaller bandwidths (see Figure \ref{fig:noisemodelingsignificance}), which then factors into $\Delta d$. This often results in values of $r$ that are overall lower for smaller bandwidths. Exceptions arise when the bandpass combination changes with changes in bandwidth. We discuss this further in Section \ref{sec:discussion-snr}.

By selecting a set of bandpasses that optimizes the differentiation between the warm Neptune and the Earth (Figure \ref{fig:colorcolor_warmneptune}), we find that Option 3a with 20\% bandpasses is the most effective at separating the warm Neptune and the Earth ($r=2$), compared to the ratios for the other options and bandpasses averaging around 1.6. For this option, the cold Neptune shows a ratio $r=2.17$. A comparable value for the cold Neptune is also achieved by option 3b, where the UV bandpass is replaced with a secondary VIS bandpass at 640 nm.
Keeping the same option but reducing the bandwidth to 10\% would marginally favor the differentiation of the cold Neptune compared to the Earth. This happens because of the variation in the VIS bandpass wavelength for the Warm Neptune case (880 nm at $\le10\%$ bandwidth, 640 nm for 20\%). Longer VIS-channel wavelengths favor the detection of the cold scenario more effectively than the warm Neptune. 

The cold Neptune is generally better differentiated compared to the warm Neptune even in the case where we maximize for warm Neptune differentiation. In other words, even when the bandpasses are optimized for the highest significance in the warm Neptune scenario, the fact that the warm Neptune spectrum is more similar to the Earth's spectrum across the wavelength range (causing an overall lower significance) would not cause an undisputed differentiability in color-color space. On the other hand, since the cold Neptune is much more differentiable from the Earth for most wavelengths, even a suboptimal bandpass choice could qualitatively separate the two models in color-color space. 

When optimizing the bandpasses for the cold Neptune (Figure \ref{fig:colorcolor_coldneptune}), we find that Options 3b at 20\% bandwidth strongly separates the Earth from a cold Neptune ($r=3.6$), followed by Option 3c at $r=2.94$. Reducing the bandwidth to 10\% would still yield promising results in the 3b case ($r=2.86$). However, the warm Neptune is not sufficiently differentiated in any combination of these bandpasses. The best scenario for this is Option 3b at 20\% bandwidth, for which $r\sim1.7$, followed by Option 3a at 20\% bandpass ($r\sim1.5$).

From this analysis, we find that the combination of 360 nm + 500 nm + 1110 nm would be the most promising to differentiate both the cold and the warm Neptune case. However, the exposure time would need to be increased to reduce the uncertainties and obtain a value of $r\ge3$; we discuss this in detail in Section \ref{sec:discussion-snr}.

\section{Discussion}\label{sec:discussion}

In this work, we assessed the potential of simultaneous multi-bandpass photometry as a key to breaking degeneracies inherent in single-bandpass, single-epoch observations of exoplanets. As shown in Sections \ref{sec:orbits} and \ref{sec:fluxes}, a single detection point can be explained by multiple orbital configurations, and planets with different compositions can produce similar fluxes in the single bandpass that HWO could use for its discovery phase (500 nm, 20\% bandwidth). While longer integration times and improvements in the instrumentation can reduce uncertainties, they cannot fully eliminate this fundamental ambiguity.

Visiting the system at multiple epochs is the primary way to assess orbital parameters, and is therefore considered as a core part of HWO’s exoplanet survey strategy and taken into account when calculating exoplanet yields for the mission \citep[see e.g.,][]{2014ApJ...795..122S,2024SPIE13092E..5MM}. Such a strategy, however, comes with relevant complications: as time passes between each visit, the orbital position changes and therefore the portion of illuminated surface, the orientation of the planet, and the atmospheric dynamics would most likely have impacted the spectrum of the planet. Hence, multi-epoch observations are not trivial to interpret, and future work is still needed to model such variations, as well as accurately constrain the atmospheric composition. This is even more complicated for multi-planetary systems, where the pairing of the photometric detection over multiple epochs with the correct planet can be challenging and requires specialized algorithms \citep[see e.g.,][]{2021SPIE11823E..0FM,2025AAS...24611604H}.

Multi-bandpass photometry during the first visit offers instead a promising approach to qualitatively constrain a planet's spectral characteristics, potentially allowing for more efficient planning and prioritization of follow-up observations. This strategy could also align well with the current design considerations for HWO and the potential presence of multiple coronagraphic channels. By using parallel channels at the same time as the detection observation, we can gain additional data at no extra observing time cost. However, broadband photometry is not meant to be the ultimate characterization method. Spectroscopy will still be necessary to provide an in-depth characterization of the target. But the additional information that multi-band photometry could provide during the first visit would allow us to provide qualitative classification that could help guide future observations, in a context where observing time is at a premium, as well as providing useful information for future visits in the case of multi-planet systems. 

\subsection{Optimizing Bandpass Choices for Planet Differentiation}

By using the Exposure Time Calculator that we developed (Section \ref{sec:noise}, \citet{pyedith2025}) we estimate the measurement uncertainty for any potential ancillary bandpass given the observing time required to achieve \snr=7 in the discovery observation (500 nm), the current baseline assumed by HWO. This allows us to identify the most promising bandpasses for various combinations of parallel channels to differentiate the Earth from ``false positives'' such as warm and cold Neptunes, thus informing an observational strategy for HWO optimized for multiple bandpasses (Section \ref{sec:observations}). 

{As shown in Table \ref{tab:bandpassesneptunes}}, the choice of optimized bandpasses strongly depends on where warm and cold Neptune spectra differ most from the Earth. 

In order to qualitatively differentiate an Earth from a cold Neptune, a secondary bandpass in the VIS range (around 850-880 nm, 20\% bandwidth) would be preferred and achieve statistical significance above the $S_{option}=3$ threshold. This is motivated by the suppressed infrared flux caused by methane and collision-induced absorption in the cold Neptune case, which makes it distinguishable from the Earth (see Figure \ref{fig:coldneptunesorbitsspectra}). If three bandpasses are available, then another bandpass in the NIR (around 1.11 micron, Option 3b at 20\% bandwidth) in addition to the second VIS bandpass would allow us to separate Earth and cold Neptune with a significance of $S_{3b,20\%}\sim4$. This result is also consistent in color-color space ($r_{3b,20\%}=3.6$, Figure \ref{fig:colorcolor_coldneptune}). A close second is Option 3c with an ultraviolet bandpass (around 300 nm, 20\% bandpass) in addition to the second VIS bandpass, whose significance is $S_{3c,20\%}=3.8$ and $r_{3c,20\%}\sim3$. 

Warm Neptune differentiation is more challenging at the baseline \snr. No bandpass combination achieves the $S_{option}=3$ threshold in significance that we require for qualitative discrimination. When only two bandpasses are available, a secondary bandpass in the UV range (Option 2a, bandpass 20\%) is preferred to reach $S_{2a,20\%}\sim 2$ sensitivity. When three bandpasses are available, an additional bandpass in the NIR range (1.11 micron, Option 3a) or another bandpass in the VIS range (around 640 nm, Option 3c) are preferred, for which the significance reaches $S_{3a,20\%}\sim S_{3c,20\%}\sim2.7$.  Option 3a is slightly favored in the color-color ratio metric ($r_{3a,20\%} =2$ compared to $r_{3c,20\%}=1.6$). From this analysis, it seems that the presence of a UV channel is necessary to discriminate the Earth and the warm Neptune, but that in general a discrimination of these two scenarios would be challenging given the broad similarities in the spectral shape. 

Even though a secondary VIS bandpass appears favorable for both the warm and the cold Neptune, these bandpasses are not placed in the same wavelength range for the two scenarios. A bandpass with a central wavelength around 850 nm maximizes the differentiation potential for the cold Neptune, but the warm Neptune has a very comparable flux to that of the Earth in that bandpass. Similarly, the bandpass centered at 640 nm which maximizes the differentiation of the warm Neptune to the Earth is at a local significance minimum for the cold Neptune case.

This fundamental trade-off prevents the bandpass combinations from being universally optimal and highlights the model-dependent nature of our findings - we should expect that different models would yield different results and that a photometric observation strategy might take into account some additional population-related statistics (see Section \ref{sec:limitations}).

Distinguishing a cold Neptune from an Earth still appears to be easier because of the larger differences in the spectra for wavelengths larger than 800 nm: even when bandpasses are optimized for the warm Neptune case (Figure \ref{fig:colorcolor_warmneptune}), the cold Neptune typically achieves $r\ge 2$. The same cannot be said for the warm Neptune case, for which optimized bandpasses are essential to get a $S_{option}\ge2$ significance level, and performance degrades when using bandpasses optimized for a cold Neptune  (Figure \ref{fig:colorcolor_coldneptune}).

Broadband filters (20\% bandwidth) are generally preferred compared to narrower ones, since they can collect more photons within the same exposure time and therefore reduce the noise of the observation. On the other hand, the flux measurement is averaged out and loses spectral information. Preferring wide bandwidths aligns with the goal to obtain a qualitative characterization during initial observations, rather than detailed spectral analysis which would be reserved for promising targets in follow-up studies. There is one notable exception: a narrower-band filter of 5\% bandwidth would leverage a maximum in differentiation significance around 1.03 $\mu$m. Wider bandpasses cannot exploit this feature due to the currently assumed boundary between VIS and NIR channels at 1 micron, which would be within the 20\% bandpass centered at 1.03 $\mu$m. This result is dependent on the the location of the boundary between the VIS and NIR channels, which is still somewhat uncertain and could be shifted towards shorter or longer wavelengths depending also on current and future detector technologies. Considering a channel edge around 930 nm would unlock the use of a filter in the high-significance region around 1.03 $\mu$m while maintaining the advantages of broadband filters. {However, this would split the 940 nm water band into two separate channels, potentially impacting characterization. We will explore the impact of the channel edge choice in a future study.}

These findings quantify the potential of multi-bandpass photometry as a triage method to prioritize further observations. However, planets that more closely resemble the Earth's spectrum (e. g., the warm Neptune) could still be confused at the baseline \snr, and this motivates further exploration of whether investing additional observing time during the first visit would alleviate the problem.

\subsection{\snr~Considerations} \label{sec:discussion-snr}

\begin{figure*}
  \centering
  \includegraphics[width=\linewidth]{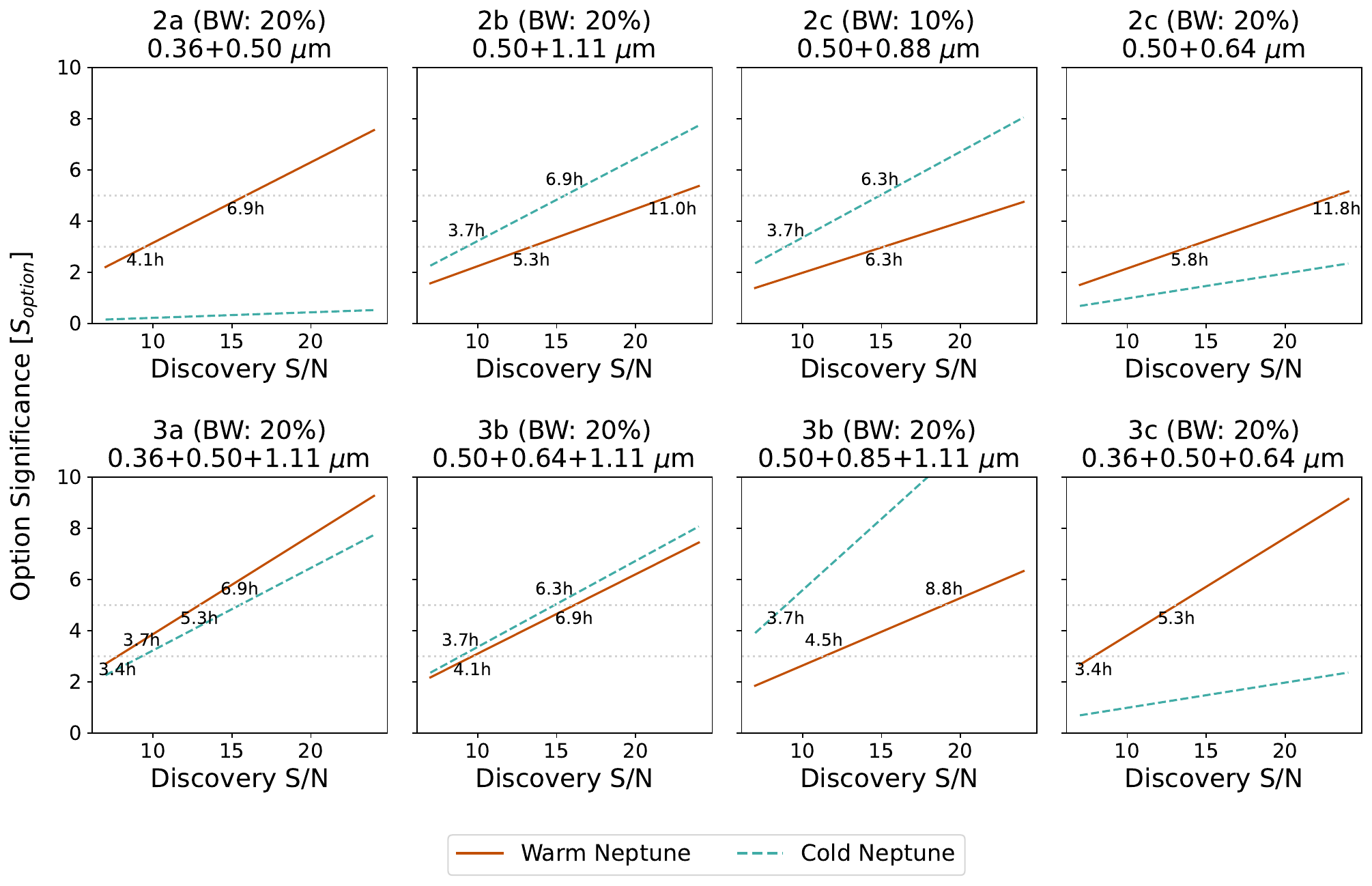}

    \caption{Total significance at varying \snr~values for {the 2- (\emph{top row}) and 3-bandpass options (\emph{bottom row}) }that were identified as promising to differentiate a warm or a cold Neptune from the Earth (see Section \ref{sec:observations}). In orange, solid lines the significance for the warm Neptune case; in aqua, dashed lines the significance for the cold Neptune case. As dashed lines, the $S_{option}=3$ and $S_{option}=5$ significance thresholds. Labeled in each subplots are the corresponding discovery exposure times (in hours) that are necessary to achieve $S_{option}=3$ and $S_{option}=5$, as calculated with \pyedith. At the top of each subset, the option keyword, the corresponding bandwidth (BW) and the center wavelengths of the bandpasses considered in the option.}
     \label{fig:highsnr}
\end{figure*}

For the bandpass combinations that were identified as promising for discriminating the Earth from either Neptune-like atmosphere at the baseline discovery \snr=7, we ran ancillary noise simulations assuming increasing discovery \snr~values. This is necessary to identify the bandpass combinations that would be best suited to qualitatively differentiate multiple planet types whenever possible, since we will not know a priori the nature of the planet when observing it for the first time. Hence, we analyzed how bandpasses that are optimized for a specific case would behave for the opposite scenario, for increasing observing time. This ancillary study provides information on the value that observing for longer exposure times has on the overall result over multiple scenarios, and allows us to prioritize specific bandpass combinations that favor both models at the same time, rather than a single scenario. We discuss future plans on how to generalize this effort in Section \ref{sec:limitations}. 

In {Figure \ref{fig:highsnr}} we show the option significance that the two- and three-bandpass options we mentioned in the previous section yield for increasing discovery \snr~to differentiate both scenarios. We also label the exposure time that would be necessary to reach $S_{option}=3$ and $S_{option}=5$ significance levels for each model, calculated with \pyedith. These times also include overheads to slew the telescope and dig the dark hole (see Section \ref{sec:noise}).

For two-bandpass observations ({top row of Figure \ref{fig:highsnr}}), a clear preference emerges for an ancillary near-infrared secondary bandpass (Option 2b, 0.5+1.1 $\mu$m, 20\% bandwidth) over alternatives (second subplot). This bandpass combination is optimized for both the warm and the cold Neptune scenario. With this bandpass combination, the cold Neptune would reach $S_{2b,20\%}=3$ after 3.7 hours (discovery \snr=9) and $S_{2b,20\%}=5$ for 7 hours (discovery \snr=15); the warm Neptune would need 5 and 11 hours to reach $S_{2b,20\%}=3$ and $S_{2b,20\%}=5$ (discovery \snr=13 and \snr=23 respectively). Observing for about 5 hours (less than twice the observing time for the currently assumed discovery \snr=7) would therefore qualitatively distinguish both scenarios to $S_{2b,20\%}=3$. 

Similar performance is obtained with a secondary VIS bandpass at 880 nm at 10\% bandwidth (third subplot), though the warm Neptune scenario would require at least 6 hours to reach the $S_{2c,10\%}=3$ level. Choosing a secondary VIS bandpass around 640 nm (fourth subplot) would yield comparable results for the warm Neptune to the ones that consider a near infrared bandwidth, while poorly differentiating the cold Neptune scenario, reducing its priority for generalized applications. Option 2a (360+550 nm, first subplot) was considered in the previous section as potentially significant to discriminate the warm Neptune, and increasing the \snr~would achieve $S_{2a,20\%}=3$ and $S_{2a,20\%}=3$ thresholds for around 4 and 7 hours respectively for this scenario. However, for this combination the cold Neptune would not be differentiable from the Earth, thus decreasing the priority of this option.

Three-bandpass observations ({bottom row of Figure \ref{fig:highsnr}}) show a general preference towards both a UV and a NIR secondary bandpass (0.36+0.5+1.11 $\mu$m, 20\% bandwidth, first subplot), for which both warm and cold Neptune would achieve significance $S_{3b,20\%}=3$ and $S_{3b,20\%}=5$ at $\sim$ 3.5 and 5-7 hours respectively. A negligible increase in exposure time (3.2 to 3.5 hours) would then already provide qualitative differentiation, with the potential of a reliable differentiation by doubling the observing time. 

Without UV coverage, multiple variants of Option 3b (VIS+Discovery+NIR) could be used, based on the choice of the VIS bandpass. For these, we label the center wavelength of the VIS bandpass as an additional subscript. By assuming a VIS bandpass at 640 nm (second subplot), which optimizes for warm Neptune differentiation, we would be able to achieve $S_{3b,20\%,640nm}=3$ significance for both planets around \snr$\sim$9 ($\sim$4 hours observing time) and $S_{3b,20\%,640nm}=5$ around \snr=15 ($\sim$6.5 hours). When we assume a VIS bandpass at longer wavelengths (optimizing for the cold Neptune scenario, third subplot), we observe that the cold Neptune is already differentiated from the Earth to $S_{3b,20\%,850nm}\ge3$ at the nominal \snr$=7$. Increasing \snr~increases the significance, reaching $S_{3b,20\%,850nm}=5$ at 3.7 hours. On the other hand, this option requires \snr=11 (4.5 hours) and \snr=19 (9 hours) to differentiate the warm Neptune scenario above the $S_{3b,20\%,850nm}=3$ and $S_{3b,20\%,850nm}=5$ thresholds.  Between these two variations of Option 3b, we would prefer the former, which would successfully differentiate both planets above $S_{3b,20\%,850nm}=5$ in 7 hours, a comparable result to Option 3a previously described. Option 3c (fourth subplot), a promising one for differentiating the warm Neptune, does not allow differentiation of the cold Neptune scenario and it is not a preferred combination.

Investing more time in single-epoch observations compared to multiple visits might prove to be useful in the observing strategy for HWO. Achieving definitive differentiation would require \snr=15-20, which would translate in 6-10 hours of observing time. While this doubles or triples the observing time compared to the baseline 3.2 hours required for \snr=7 (see Section \ref{sec:noise}), it is still substantially less than spectroscopic alternatives and avoids the orbital confusion that appears in multi-epoch strategies. It would also be a time-saver in terms of overheads: slew, dark hole digging, thermal stabilization upon pointing would only be performed once for a single epoch, while they would need to be repeated for multiple visits. However,  these findings will need be further assessed in the context of the overall yield of the mission (see Section \ref{sec:limitations}).

\subsection{Comparing with Previous Work} \label{sec:discussion-comparison}
\citet{JKT2016} answered a similar question to what we aim to do in this study. Through maximization of the distance between the Earth and any other planet in color-color space, the authors could identify combinations of bandpasses that would allow differentiation of the Earth from false positives and found optimized bandpasses around 431-531 nm, 569-693 nm, 770-894 nm. These values stem from the optimization algorithm that they used, spanning all possible wavelength combinations and do not take into account any noise associated with the measured flux to identify the optimized bandpasses. Rule-of-thumb calculations on the desired noise level to differentiate the Earth from other false positives was considered at a later stage, where the authors discussed the value of photometry compared to spectroscopy in detector-dominated regimes. While this is outside the scope of this work (see Section \ref{sec:limitations}),  a critical advancement in our approach is the integration of realistic noise modeling throughout the bandpass optimization process. 

The color-color space distance is a metric that we also consider in this paper (see Section \ref{sec:observations}), although we identify the best bandpasses  based on their differentiation potential, considering the expected noise for each photometric band (see Section \ref{sec:noise}). We therefore include the noise at an earlier stage in the process and that informs the bandpass optimization directly. This can be seen for example in the near-infrared, where the significance decreases as the noise increases and no bandpasses are chosen at longer wavelengths, although noticeable differences between the Earth and the cold Neptune could be identified (see Figure \ref{fig:coldneptunesorbitsspectra}).

\citet{JKT2016} considered a wider range of planets for their calculation, though phase was mostly kept fixed at 60$^\circ$ except for some cases. In our work, we consider phases that could be reasonably plausible with a given detection (satisfying the orbital condition, see Section \ref{sec:orbits}). Results are however highly dependent on phase, and we should expect variations when assuming an Earth-like planet not at quadrature, as was highlighted in \citet{JKT2016}.

The authors also focused on the VIS range only, while we expanded to the ultraviolet and near-infrared; they also set the width of the bandpasses to be at least 100 nm, which would make them similar to our 20\% bandwidth case. Nevertheless, the VIS bandpasses that were identified in that study are overall similar to the ones we identified in our work.

\subsection{Limitations and Future Work} \label{sec:limitations}

This study is a first exploration of the potential for multi-bandpass photometric planet discrimination with HWO, establishing foundational methodology while making several simplifying assumptions.

We focused on differentiating the Earth from (warm and cold) Neptune-like planets since it was motivated by previous work showing how these planet archetypes could show similar features \citep{2018haex.bookE..98R}. Even in this simplified case, we could already demonstrate how bandpass optimization is extremely dependent on specific planetary spectral features. We can imagine that, extending the study over multiple planetary classes covering a variety of potential atmospheres, would likely yield different bandpass optimizations. We also did not consider planet occurrence rates in our analysis, though they might play a role in determining how likely it might be that an Earth-like planet might be confused with a Neptune on a specific orbit. Occurrence rates for HZ terrestrial planets around Solar-like stars (as well as larger planets in more distant orbits) are still lacking, and are expected to be one of the main outcomes of upcoming missions such as PLATO \citep[PLAnetary Transits and Oscillations of stars][]{Rauer2024-PLATOMission}. Studies like this could be periodically carried out as these estimates are refined.

We employed fixed templates for the warm and cold Neptune, allowing the flux to only be impacted by orbital parameters and focusing our attention on the effects of orbital geometry on discrimination performance. We expect, however, that the atmosphere of a planet would also be impacted by its orbital features: the instellation received varies dramatically with distance from the star and that would impact both the thermal profile and the atmospheric chemistry. A more accurate approach would be to calculate a grid of models that takes into account variations in atmospheric profiles, although an accurate computation would require a larger effort and substantial computing power. The nature of such atmospheres is also predicted to be very different: gaseous planets span a range of cloud pressures and methane and water vapor mixing ratios above the cloud deck, all of which impacts the spectral features and its shape. The radius could also be varied, for example smaller radii for warm Neptunes which are predicted to be abundant and likely to cause confusion.
Furthermore, we have not considered variations in the spectrum of the Earth during its evolution, nor the broader range of terrestrial atmospheres that have been explored in earlier studies \citep{JKT2016,Smith2020}. Future studies are foreseen to explore different planetary classes to provide statistically robust bandpass combinations that could separate habitable planets from false positives. This could be done with the methodology developed in this study, which is fully independent from the underlying physics that shape the spectra.

In this work, we assumed circular orbits, prescribing a planet detected at a specific projected separation. In reality, orbits can be eccentric and their orientation can also vary ($\Omega$ is not fixed at $0^\circ$). The effect of this assumption is negligible in this work, since we are dealing with a single epoch and the apparent phase only depends from the angle between the star and the planet, and the one between the planet and the observer (see Section \ref{sec:orbits}). However, this would play a decisive role when considering multiple epochs or when considering the climate evolution of planets in eccentric orbits. We also assumed that no other known planets are in the system; in reality, a known wider companion might provide information on the likely orientation of the orbital plane, which could reduce orbital degeneracies and improve discrimination strategies. In a future publication we expect to address the added benefit of multi-bandpass photometry as a qualitative discrimination method to correctly associate observations of multi-planetary systems performed at different epochs, which could simplify the orbit disentanglement.

We also only considered a Sun-{twin} star, a class that is included in the putative HWO sample but that does not define the entirety of scenarios that could be analyzed. Different stellar classes would impact the observations both in terms of planetary flux (through different levels of irradiation and the impact of photochemistry) and noise features (through variations in stellar leakage and zodiacal background). 

Concerning technical assumptions, we arbitrarily assumed a 10\% uncertainty on the measurement of the projected separation, though this measurement is dependent on detector pixel size and post-processing techniques. Increasing the uncertainty on the separation would increase the number of potentially compatible orbits and subsequently the number of orbital configurations and planets that could reasonably cause confusion. However, this does not change the methodology itself, only the range of scenarios that would cause confusion.

We also assumed idealized rectangular filter profiles as these still have yet to be defined, but we expect real filters to have wavelength-dependent throughput and the efficacy of a specific observing option would depend on the current availability of filters and detectors. We considered an idealized concept for HWO, which in some ways resembles the first Exploratory Analytic Case that has been developed by the HWO Project Office, but that does not capture wavelength-dependent effects. This is still a useful exercise as the architecture trade studies progress, and provides generalized results and access to a broad parameter space, at the expense of high fidelity towards a specific concept. We would therefore expect that other potential bandpass combinations may be preferred once such parameters are assessed as the project reaches a more definitive design.

We did not discuss in detail the benefit of photometry compared to spectroscopy, focusing on pure photometric optimization. On the other hand, the marginal performance for the warm Neptune raised the question whether there could be optimal combinations of photometric and spectroscopic observations to enhance characterization. This will be explored in a dedicated study. 

The most optimistic observing scenarios capable of distinguishing between the two Neptune cases involve three bandpasses (configurations 3a or 3b), all observed in parallel. However, instrument constraints or triaging across diverse planetary scenarios may require sequential follow-up observations. In such cases, an observational decision tree becomes a critical tool for guiding the sequencing of follow-up observations and prioritizing targets. Trade studies exploring combinations like two parallel bandpasses followed by a targeted observation, or two sequential sets of parallel bandpasses that enhance spectral coverage would benefit from being framed within a decision tree approach. This framework allows initial discovery data to inform whether additional observations are needed to achieve sufficient confidence for final categorization. It can also help optimize efficiency by minimizing total observing time and scheduling overhead. While providing mission-level constraints and trade-offs for potential HWO observing strategies based on multi-bandpass photometry was beyond our current scope, our methodology also provides the foundational photometric discrimination capabilities to be integrated in HWO's yield models and target prioritization algorithms.

\section{Summary} \label{sec:summary}

In this work, we explored the potential of multi-bandpass photometry as a technique to employ during the detection of a planet with HWO. By using multiple photometric bandpasses in addition to the discovery one, we identify promising bandpass combinations that could differentiate an Earth-like planet from Neptune-like ones. This enables quicker target prioritization and subsequent optimization of the mission scheduling.

We developed a comprehensive {open-source} algorithm\footnote{{Available at: \href{https://github.com/eleonoraalei/mpear}{https://github.com/eleonoraalei/mpear}} }  that explores the parameter space and identifies potentially confusing scenarios, both in terms of orbital parameters and flux in the discovery bandpass {\citep{alei_2026_18258482} }. We modeled orbits of potential (cold and warm) Neptune-like planets that would be consistent with the same on-sky projected distance as the Earth at quadrature orbiting a Sun-{twin} star at 10 pc. We then modeled the corresponding spectra and found the scenarios that would yield a comparable flux at 500 nm (the discovery bandpass assumed by HWO), thus making the Earth and the Neptune-like planets virtually impossible to distinguish with a single photometric data point. We then evaluated the noise that any other potential bandpass across the wavelength range would have for the exposure time determined by the discovery observation (to achieve \snr=7, approximately 3.2 hours). To achieve this, we used the HWO coronagraph exposure time calculator that we developed, which we call \pyedith. We then identified the  2- and 3-bandpass combinations that would yield the largest statistical significance in differentiating between the spectrum of an Earth and those of warm and cold Neptune-like planets, i. e., the largest differentiation potential to distinguish an Earth from a Neptune. 
We find that:
\begin{itemize}
\item Only a selected number of orbits can intersect a given projected separation. This reduces the orbital configurations that can match the detection. Considering that HWO will acquire a photometric flux measurement at 500 nm, the intersection of plausible orbital configurations that would not only be consistent with the projected separation but also with the photometric flux shrinks even further because of the high phase-dependency of {the }reflected-light spectrum.
    \item At the baseline discovery \snr=7, there is no solution that reliably separates both a warm or a cold Neptune from the Earth, with all combinations falling short of the $S=5$ level significance. The warm Neptune cannot be differentiated with 2- or 3-bandpass options to more than $S\sim2.7$, while a second VIS bandpass and potentially a third one in the UV or NIR could qualitatively differentiate the Earth from a cold Neptune ($S=3-3.5$). \item Even if a second bandpass in the VIS range would enhance the discrimination between false positives and the Earth, the warm Neptune would be better differentiated with a VIS-channel bandpass at 640 nm while the cold Neptune differentiation would benefit from a bandpass at 850 nm. 
    \item Similar significance for both models is obtained by adding a 20\%-wide near-infrared bandpass around 1.11 $\mu$m to the discovery bandpass (Option 2b), though longer exposure times  would be needed to achieve $S_{2b,20\%}=3$ ($\sim$5 hours) and $S_{2b,20\%}=5$ (11 hours).
    \item If three bandpasses are available, the optimal combination includes UV (360 nm), discovery (500 nm) and near-infrared (1.11 $\mu$m), achieving qualitative and reliable discrimination with slightly longer observation times (3.7-7 hours for $S_{3a,20\%}=3$ and $S_{3a,20\%}=5$ levels). In the absence of the UV channel (Option 3b), a secondary VIS bandpass at 640 nm would provide confident results for both scenarios after 4 hours ($S_{3b,20\%}=3$) or 7 hours ($S_{3b,20\%}=5$).
    \item Strategic time investment during first epochs (6-7 hours compared to 3.2 hours baseline, corresponding to a discovery \snr=15 rather than the currently-assumed \snr=7) can achieve a more reliable $S_{option}=5$ discrimination between the planetary types we examined here. Future studies are needed to understand how this extra time commitment during the discovery phase changes the mission yield. 
    \item A spectral region at 1.03 $\mu$m with strong potential for spectral discrimination could not be accessed due to the assumed VIS/NIR channel boundary at 1.0 $\mu$m. Shifting the boundary to $\sim$930 nm could unlock significant discrimination improvements while allowing multiple bandpasses in each channel.

\end{itemize}

These results demonstrate that multi-bandpass photometry is a promising triage method for HWO, allowing us to start prioritizing follow-up observations from the first visit, potentially avoiding time-intensive observations when not necessary and reserving detailed characterization only for promising targets. This could significantly enhance the mission yield through more optimized use of observing time.

\begin{acknowledgments}
The authors thank Rhonda Morgan, Samantha Hasler, Sarah Steiger, Natasha Latouf, and Corey Spohn for helpful discussions. The work of  E.A. and M.H.C. was supported by appointments to the NASA Postdoctoral Program at the NASA Goddard Space Flight Center, administered by Oak Ridge Associated Universities under contract with NASA  (ORAU-80HQTR21CA005). A.M.M. and A.V.Y acknowledge support from the GSFC Sellers Exoplanet Environments Collaboration (SEEC), which is supported by NASA's Planetary Science Division's Research Program. A.M.M. acknowledges support from the Explanet Spectroscopy Technology ISFM, which is supported by NASA's Astrophysics Science Division. 

\end{acknowledgments}
\software{MPEAR \citep{alei_2026_18258482}, pyEDITH \citep{alei_2025_17917472}, spectres \citep{2017arXiv170505165C}, Astropy \citep{astropy:2013, astropy:2018, astropy:2022}, numpy \citep{harris2020array}, pandas \citep{reback2020pandas, mckinney-proc-scipy-2010}, matplotlib \citep{Hunter:2007}.}

\begin{contribution}
E.A. developed the MPEAR algorithm {(released as open-source software \citep{alei_2026_18258482})}, developed and performed the necessary pyEDITH software enhancements, performed the analysis, wrote and edited the manuscript. 
E.A., A.M.M., and A.R. conceptualized the project. A.M.M., A.R. and C.C.S. provided advice.  M.H.C. co-developed pyEDITH. A.P., V.K., G.L.V., and R.H. provided input templates. V.K., A.P., and A.V.Y. wrote portions of the manuscript. All listed authors reviewed and edited the manuscript.

\end{contribution}

\bibliography{new.ms}{}
\bibliographystyle{aasjournalv7}

\end{document}